\documentclass[aps,reprint,superscriptaddress]{revtex4-1}

\usepackage{graphicx}
\usepackage{amsmath}
\usepackage{latexsym}

\begin{document}


\title{Variational approach to time-dependent fluorescence of a driven qubit}


\author{Yiying Yan}\email{yiyingyan@zust.edu.cn}
\affiliation{Department of Physics, School of Science, Zhejiang University of Science and Technology, Hangzhou 310023, China}
\author{Lipeng Chen}
\affiliation{Laboratory of Theoretical Physical Chemistry, Institut des Sciences et Ing\'{e}nierie Chimiques, Ecole Polytechnique F\'{e}d\'{e}rale de Lausanne (EPFL), CH-1015 Lausanne, Switzerland}
\author{JunYan Luo}
\affiliation{Department of Physics, School of Science, Zhejiang University of Science and Technology, Hangzhou 310023, China}
\author{Yang Zhao}\email{YZhao@ntu.edu.sg}
\affiliation{Division of Materials Science, Nanyang Technological University, Singapore 639798, Singapore}


\date{\today}

\begin{abstract}
We employ the Dirac-Frenkel variational principle and multiple Davydov ansatz to study time-dependent fluorescence spectra of a driven qubit in the weak- to strong qubit-reservoir coupling regimes, where both the Rabi frequency and spontaneous decay rate are comparable to the transition frequency of the qubit. Our method agrees well with the time-local master-equation approach in the weak-coupling regime, and offers a flexible way to compute the spectra from the bosonic dynamics instead of two-time correlation functions. While the perturbative master equation breaks down in the strong-coupling regime, our method actually becomes more accurate due to the use of bosonic coherent states under certain conditions.  We show that the counter-rotating coupling between the qubit and the reservoir has considerable contributions to the photon number dynamics and the spectra under strong driving conditions even though the coupling is moderately weak. The time-dependent spectra are found to be generally asymmetric, a feature that is derived from photon number dynamics. In addition, it is shown that the spectral profiles can be dramatically different from the Mollow triplet due to strong dissipation and/or multiphoton processes associated with the strong driving. Our formalism provides a unique perspective to interpret time-dependent spectra.
\end{abstract}


\maketitle

\section{introduction}
Light-matter interactions play a fundamental role in quantum optics and quantum control~\cite{scully}. In recent years, light-matter interactions have attracted much attention in the strong-coupling regime where the coupling strength is comparable with the transition frequencies of the system~\cite{nori,Fornrmp,daniele,Simone,Jaako,Zueco,Paraonu}. In general, a two-level system strongly coupled to a single harmonic oscillator, as described by quantum Rabi model, is widely studied theoretically and explored experimentally in the context of artificial atoms such as superconducting circuits~\cite{FornBS,Yoshihara,Yoshihara2,langford}. It has been demonstrated in laboratory that
strong coupling between a qubit and an electromagnetic continuum (i.e., a collection of harmonic oscillators) can also be realized, whereby the spontaneous decay rate becomes comparable to or even exceeds the transition frequency of the qubit~\cite{Forn}, a situation that is dramatically different from the case of natural atoms in the free space. A natural atom interacts with an electromagnetic field in the vacuum in the weak-coupling regime, leading to spontaneous decay rates far smaller than the transition frequencies~\cite{scully}. It is therefore interesting to reconsider the elementary processes such as absorption, emission, and photon scattering in the strong-coupling regime.

Of particular interest and importance is the scattering of photons from a two-level system coherently driven by a monochromatic laser field, which gives rise to resonance fluorescence~\cite{scully}. Several features of this light scattering process have been elucidated such as the Mollow triplet~\cite{mollow}, antibunching~\cite{kimble}, and squeezing~\cite{Carmichael}. The Mollow triplet is referred to as a three-peaked fluorescence spectrum appearing when the Rabi frequency is much larger than the spontaneous decay rate~\cite{mollow}. The modification to the Mollow triplet is illustrated by taking into account of various influences, e.g., the squeezed vacuum~\cite{Toyli}, the phonon bath~\cite{Ulrich,nazir}, strong harmonic driving~\cite{Browne,zheng}, etc. In addition, the time evolution of the fluorescence is also analyzed by considering the time-dependent spectrum~\cite{Renaud,Eberly}. However, all these studies of the fluorescence spectrum are in the weak light-mater coupling regime where the spontaneous decay rate is far smaller than the transition frequency.

Although the fluorescence spectrum is related to the number of scattered photons,
it is typically calculated from a two-time correlation function of the emitter~\cite{corrrwa}.
One method is based on the quantum optical master equation and Markovian quantum regression theory~\cite{mollow}.
An extended approach is the non-Markovian quantum regression theory employing the Nakajima-Zwanzig projection technique~\cite{mccutcheon}. In addition, there is a resolvent operator formalism for fluorescence spectrum calculations \cite{harman}. In general, these approaches require a perturbation expansion in the system-reservoir coupling strength, and preferred expansions are usually done up to the second order~\cite{mccutcheon,harman,goan,Breuer}. Consequently, these methods have difficulty when applied to the strong-coupling regime. To overcome the difficulties with the master-equation approaches, a flexible way to obtain the spectrum is to directly gauge the photonic (bosonic) dynamics instead of evaluating the two-time correlation function for emitter operators. Numerical simulation of fluorescence spectra based on bosonic observable has been carried out by the means of time-dependent
density matrix renormalization group exploring the suppression of spectral diffusion with optical pulses~\cite{Fotso}. More recently, the reservoir information in open quantum system has been studied by the dynamical polaron ansatz~\cite{diaz}, the stochastic c-number Langevin equation~\cite{Zhou}, and short iterative Lanczos method~\cite{Filippis}. However, so far there is no direct evaluation of fluorescence spectrum from the bosonic degrees of freedom in the strong-coupling regime.

In this paper, we combine the Dirac-Frenkel variational principle~\cite{dfvp} with the multiple Davydov ansatz~\cite{wanglu,tssbm} to study the time-dependent fluorescence spectrum of a coherently driven qubit in the moderately weak to the strong coupling regime. The variational approach has been applied to explore the reduced dynamics and the bosonic dynamics of the spin-boson model~\cite{wanglu,tssbm,yuta,hzk}. In contrast with the master-equation approach, our method retains full information on the bosonic degrees of freedom in our equations of motion, thereby providing  simultaneous access to not only the reduced dynamics but also the photon number dynamics. In the moderately weak coupling regime, it is found that the results calculated from the variational approach are consistent with those of the master-equation approach. In the strong coupling regime, the variational results are more robust while those from master-equation approach are invalid because of the breakdown of the perturbative master equation in the absence of adequately strong driving. Moreover, the effects of counter-rotating coupling between the qubit and the reservoir on the photon number dynamics is illustrated in moderately weak coupling regimes. It is shown that the time-dependent fluorescence spectrum is generally asymmetric, which can be intuitively understood by considering photon number dynamics. In addition, we demonstrate that the spectral profiles can be dramatically distinct from the Mollow triplet in the presence of strong dissipation and/or multiphoton processes associated with the strong driving. The present formalism provides a unique perspective on the time-dependent spectrum from the bosonic degrees of freedom.

The rest of paper is organized as follows. In Sec.~\ref{sec:theory} we introduce the Dirac-Frenkel time dependent variational principle and the multiple Davydov trial state. We also describe the time-local master equation approach, which is used to evaluate the photon number in the reservoir for both original and rotating-wave approximation (RWA) Hamiltonian. In Sec.~\ref{sec:dyn}, we discuss the validity of the variational approach by comparing the qubit dynamics with those from other methods and by calculating the ansatz deviation. In Sec.~\ref{sec:result} we present the main results concerning the reservoir photon number dynamics calculated from the variational approach and master-equation approach. We compare results from the two approaches and examine the features of time-dependent spectra as we go from the moderately weak to the strong coupling regime. In Sec.~\ref{sec:con}, the conclusions are drawn.

\section{Model and methodologies}\label{sec:theory}

We consider that a qubit is subjected to a harmonic driving and
coupled with a radiative reservoir. The total Hamiltonian reads (we
set $\hbar=1$ throughout this paper)
\begin{equation}
H(t)=H_{{\rm S}}(t)+H_{{\rm R}}+H_{{\rm SR}},\label{eq:Ham}
\end{equation}
where $H_{{\rm S}}(t)$ describes the driven qubit:
\begin{equation}
H_{{\rm S}}(t)=\frac{1}{2}\omega_{0}\sigma_{z}+\Omega\cos(\omega_{x}t)\sigma_{x}
\end{equation}
$\omega_{0}$ is the transition frequency between the two levels of
the qubit and $\sigma_{z(x,y)}$ are the Pauli matrices. $\Omega$ is
the Rabi frequency and $\omega_{x}$ is the driving frequency. $H_R$ is the
reservoir Hamiltonian and is given by
\begin{equation}
H_{{\rm R}}=\sum_{k}\omega_{k}b_{k}^{\dagger}b_{k}
\end{equation}
with $b_{k}$ $(b_{k}^{\dagger})$ the annihilation (creation) operator
of the $k$th bosonic mode. $H_{{\rm SR}}$ describes the interaction
between the qubit and the reservoir and takes the form
\begin{equation}
H_{{\rm SR}}=\frac{\sigma_{x}}{2}\sum_{k}\lambda_{k}(b_{k}+b_{k}^{\dagger}),\label{eq:HSR}
\end{equation}
where $\lambda_{k}$ is the coupling strength between the $k$th mode
and qubit. In this work, we consider that the interaction between the qubit and reservoir is characterized
by the Ohmic spectral density
\begin{equation}
J(\omega)=\sum_{k}\lambda_{k}^{2}\delta(\omega-\omega_{k})=2\alpha\omega\Theta(\omega_{c}-\omega),
\end{equation}
where $\alpha$ is a dimensionless coupling strength, $\omega_{c}$
is the cut-off frequency, and $\Theta(\cdot)$ is the Heaviside step
function. A qubit ultrastrongly interacts with an Ohmic bath can be physically realized by a superconducting flux qubit coupled to an open 1D transmission line~\cite{Forn}. In addition, the present method is applicable to the Lorentzian and sub-Ohmic spectral density functions as well.

In what follows, we use Dirac-Frenkel time-dependent variational principle and multiple Davydov ansatz to calculate the dynamics of the qubit and reservoir simultaneously. Particularly, we will mainly focus on the reservoir observable related to time-dependent fluorescence spectrum.

\subsection{Dirac-Frenkel time-dependent variational
principle and multiple Davydov ansatz}

The solutions to the time-dependent Schr\"{o}dinger equation associated with the Hamiltonian (\ref{eq:Ham}) can be derived from the Dirac-Frenkel time-dependent variational principle, of which the action is defined as~\cite{dfvp}
\begin{equation}
  {\cal S}[\psi]=\int^{t_2}_{t_1}\langle\psi(t)|H(t)-i\partial_t|\psi(t)\rangle dt,
\end{equation}
where $|\psi(t)\rangle$ are trial states. The optimal solutions are obtained by restricting $\delta {\cal S}[\psi]=0$, which is equivalent to
\begin{equation}
\langle\delta \psi(t)|H(t)-i\partial_{t}|\psi(t)\rangle=0.~\label{eq:dftdvp}
\end{equation}

We will use the multiple Davydov D$_1$ trial state, also known as the multi-D$_1$ ansatz, in our variational approach, which takes the form~\cite{wanglu,tssbm}
\begin{equation}
|D_{M}(t)\rangle=\sum_{n=1}^{M}\left[A_{n}(t)|+\rangle|f_{n}(t)\rangle+B_{n}(t)|-\rangle|g_{n}(t)\rangle\right],
\end{equation}
where $|\pm\rangle$ are the eigenstates of $\sigma_{x}$, $|f_{n}(t)\rangle$
and $|g_{n}(t)\rangle$ are multimode Bargmann coherent states (which differ from the Glauber's coherent states by a normalization factor)~\cite{Werther}:
\begin{equation}
|f_{n}(t)\rangle=\exp\left[\sum_{k}f_{nk}(t)b_{k}^{\dagger}\right]|\{0_{k}\}\rangle,
\end{equation}
\begin{equation}
|g_{n}(t)\rangle=\exp\left[\sum_{k}g_{nk}(t)b_{k}^{\dagger}\right]|\{0_{k}\}\rangle,
\end{equation}
with $|\{0_{k}\}\rangle$ being the vacuum state of the reservoir.
$M$ is the multiplicity of the Davydov trial state. Specifically,
when $M=1$, the trial state reduces to the single Davydov state.
In the above ansatz, we have introduced a set of variational parameters:
$A_{n}(t)$, $B_{n}(t)$, $f_{nk}(t)$, and $g_{nk}(t)$. The physical
significance of these parameters are clear: $A_{n}(t)$ and $B_{n}(t)$
are the probability amplitudes while $f_{nk}(t)$ and $g_{nk}(t)$
are the displacements of the $k$th mode. For the sake of simplicity, we shall use the simplified notations: $A_n\equiv A_n(t)$, $B_n\equiv B_n(t)$, $f_n\equiv f_n(t)$, and $g_n\equiv g_n(t)$ hereafter. The equations of motion
for the variational parameters are determined by substituting the ansatz into Eq.~(\ref{eq:dftdvp}) and are given as follows:
\begin{widetext}
\begin{eqnarray}
0 & = & -i\sum_{n=1}^{M}\left(\dot{A}_{n}+A_{n}\sum_{k}\dot{f}_{nk}f_{lk}^{\ast}\right)S_{ln}^{(f,f)}+\sum_{n=1}^{M}\frac{\omega_{0}}{2}B_{n}S_{ln}^{(f,g)}\nonumber \\
 &  & +\sum_{n=1}^{M}A_{n}\left[\sum_{k}\omega_{k}f_{lk}^{\ast}f_{nk}+\Omega\cos(\omega_{x}t)+\sum_{k}\frac{\lambda_{k}}{2}(f_{lk}^{\ast}+f_{nk})\right]S_{ln}^{(f,f)},\label{eq:Aeq}
\end{eqnarray}
\begin{eqnarray}
0 & = & -i\sum_{n=1}^{M}\left(\dot{B}_{n}+B_{n}\sum_{k}\dot{g}_{nk}g_{lk}^{\ast}\right)S_{ln}^{(g,g)}+\sum_{n=1}^{M}\frac{\omega_{0}}{2}A_{n}S_{ln}^{(g,f)}\nonumber \\
 &  & +\sum_{n=1}^{M}B_{n}\left[\sum_{k}\omega_{k}g_{lk}^{\ast}g_{nk}-\Omega\cos(\omega_{x}t)-\sum_{k}\frac{\lambda_{k}}{2}(g_{lk}^{\ast}+g_{nk})\right]S_{ln}^{(g,g)},\label{eq:Beq}
\end{eqnarray}
\begin{eqnarray}
0 & = & -i\sum_{n=1}^{M}\left[\dot{A}_{n}f_{np}+A_{n}\sum_{k}\dot{f}_{nk}(\delta_{p,k}+f_{lk}^{\ast}f_{np})\right]S_{ln}^{(f,f)}+\sum_{n=1}^{M}\frac{\omega_{0}}{2}B_{n}g_{np}S_{ln}^{(f,g)}\nonumber \\
 &  & +\sum_{n=1}^{M}A_{n}\left[\sum_{k}\omega_{k}(\delta_{p,k}+f_{lk}^{\ast}f_{np})f_{nk}+\Omega\cos(\omega_{x}t)f_{np}+\frac{\lambda_{p}}{2}+\sum_{k}\frac{\lambda_{k}}{2}(f_{lk}^{\ast}+f_{nk})f_{np}\right]S_{ln}^{(f,f)},\label{eq:Feq}
\end{eqnarray}
\begin{eqnarray}
0 & = & -i\sum_{n=1}^{M}\left[\dot{B}_{n}g_{np}+B_{n}\sum_{k}\dot{g}_{nk}(\delta_{p,k}+g_{lk}^{\ast}g_{np})\right]S_{ln}^{(g,g)}+\sum_{n=1}^{M}\frac{\omega_{0}}{2}A_{n}f_{np}S_{ln}^{(g,f)}\nonumber \\
 &  & +\sum_{n=1}^{M}B_{n}\left[\sum_{k}\omega_{k}(\delta_{p,k}+g_{lk}^{\ast}g_{np})g_{nk}-\Omega\cos(\omega_{x}t)g_{np}-\frac{\lambda_{p}}{2}-\sum_{k}\frac{\lambda_{k}}{2}(g_{lk}^{\ast}+g_{nk})g_{np}\right]S_{ln}^{(g,g)},\label{eq:Geq}
\end{eqnarray}
\end{widetext}where
\begin{equation}
S_{ln}^{(f,g)}=\langle f_{l}(t)|g_{n}(t)\rangle=\exp\left[\sum_{k}f_{lk}^{\ast}g_{nk}\right].
\end{equation}
The detailed derivation is presented in Appendix~\ref{sec:tdvp}.

The equations of motion (\ref{eq:Aeq})-(\ref{eq:Geq}) can be solved
numerically via the 4th-order Runge-Kutta method.
Note that the equations of motion represent a set of linear equations, which can be written in a matrix from ${\cal M}\vec{\dot{y}}=\vec{b}$, where ${\cal M}$ denotes the coefficient matrix, $\vec{\dot{y}}$ denotes the vector composed of the derivatives of the variational parameters, and $\vec{b}$ the inhomogeneous term. By solving the linear equations, one obtains the derivatives of the variational parameters,  which are used to calculate the values of the variational parameters at later times with the Runge-Kutta method.

To perform numerical simulation, we specify
$\lambda_{k}$ and $\omega_{k}$ by using the linear discretization
of the spectral density. We divide the frequency domain $[0,\omega_{c}]$
into $N_{b}$ equal intervals $[x_{k-1},x_{k}]$ with $x_{k}=k\omega_{c}/N_{b}$
($k=0,1,2,\ldots,N_{b}$). The coupling strength and frequency
for the $k$th mode are given as
\begin{equation}
\lambda_{k}^{2}=\int_{x_{k-1}}^{x_{k}}J(\omega)d\omega,
\end{equation}
\begin{equation}
\omega_{k}=\lambda_{k}^{-2}\int_{x_{k-1}}^{x_{k}}\omega J(\omega)d\omega.
\end{equation}
In addition, we specify the initial state of the total system. We assume that the qubit and reservoir are initially in a factorized state (this assumption is not necessary), where the qubit may be in the
excited state $[A_{n}(0)=B_{n}(0)=\delta_{n,1}/\sqrt{2}]$ or ground state $[A_{n}(0)=-B_{n}(0)=\delta_{n,1}/\sqrt{2}]$ while the
reservoir is in the vacuum state $[f_{nk}(0)=g_{nk}(0)=0]$.

On numerically solving the equations of motion, we obtain $A_{n}(t)$,
$B_{n}(t)$, $f_{nk}(t)$, and $g_{nk}(t)$ and thus can calculate
physical quantities of interest. Particularly, the number of photon
in the $k$th mode at time $t$ can be directly obtained:

\begin{eqnarray}
N(\omega_{k},t) & = & \langle D_{M}(t)|b_{k}^{\dagger}b_{k}|D_{M}(t)\rangle\nonumber \\
 & = & \sum_{n,l=1}^{M}\left[A_{l}^{\ast}f_{lk}^{\ast}S_{ln}^{(f,f)}f_{nk}A_{n}+B_{l}^{\ast}g_{lk}^{\ast}S_{ln}^{(g,g)}g_{nk}B_{n}\right],\nonumber\\
\end{eqnarray}
Physically, $N(\omega_{k},t)$ counts the number of photon
scattered into the $k$th mode of the initially vacuum reservoir. Thus $N(\omega_k,t)$ as a function of $\omega_k$ and $t$ can be regarded as the time-dependent
fluorescence spectrum. This is an advantage of the present method,
which directly retains the degrees of freedom of the reservoir that are traced
out in the master-equation approaches.

The reduced dynamics of the
qubit such as the population difference of the qubit can also be calculated
as
\begin{eqnarray}
P_{z}(t) & = & \langle D_{M}(t)|\sigma_{z}|D_{M}(t)\rangle\nonumber \\
 & = & \sum_{n,l=1}^{M}\left[A_{l}^{\ast}S_{ln}^{(f,g)}B_{n}+B_{l}^{\ast}S_{ln}^{(g,f)}A_{n}\right].
\end{eqnarray}
In addition, the norm of the multi-D$_1$ ansatz can be determined via
\begin{eqnarray}
{\cal N} & = & \sqrt{\langle D_{M}(t)|D_{M}(t)\rangle}\nonumber \\
 & = & \left\{\sum_{n,l=1}^{M}\left[A_{l}^{\ast}S_{ln}^{(f,f)}A_{n}+B_{l}^{\ast}S_{ln}^{(g,g)}B_{n}\right]\right\}^{\frac{1}{2}}.
\end{eqnarray}
The norm should be equal to $1$ within the evolution time of interest if the numerical solutions are convergent and the initial state is normalized. Convergence and accuracy of the variational approach will be discussed in Sec.~\ref{sec:dyn}.

\subsection{Time-local master equation approach}

\subsubsection{Reservoir photon number evaluated from a two-time correlation function}

Master equations are widely used to describe the reduced dynamics
of open quantum systems, after tracing out the degrees of freedom of
the reservoir. Nevertheless, the master-equation approach can also
calculate the photon number $N(\omega_{k},t)$ in the $k$th mode at the expense of evaluating a two-time
correlation function. Such a correlation function can be derived in
the Heisenberg picture, where the annihilation operator becomes time
dependent and is given by

\begin{equation}
b_{k}(t)=e^{-i\omega_{k}t}b_{k}(0)-i\frac{\lambda_{k}}{2}\int_{0}^{t}\sigma_{x}(t_{1})e^{-i\omega_{k}(t-t_{1})}dt_{1},
\end{equation}
where $\sigma_{x}(t)=U^\dagger(t)\sigma_{x}U(t)$ is the Pauli matrix in the Heisenberg picture. Here $U(t)$ is the unitary evolution operator for the total Hamiltonian.
This relation leads to
\begin{eqnarray}
N(\omega_{k},t)&=&{\rm Tr}[b^\dagger_{k}(t)b_{k}(t)\rho(0)]\nonumber\\
 &=&\frac{\lambda_{k}^{2}}{4}\int_{0}^{t}\int_{0}^{t}\langle\sigma_{x}(t_{1})\sigma_{x}(t_{2})\rangle e^{-i\omega_{k}(t_{1}-t_{2})}dt_{1}dt_{2},\nonumber\\\label{eq:sxsx}
\end{eqnarray}
where $\langle\cdot\rangle$ denotes the average with respect to the
initial state of the qubit and reservoir $\rho(0)=\rho_{{\rm S}}(0)\otimes|\{0_{k}\}\rangle\langle\{0_{k}\}|$, which is a direct product of the qubit state $\rho_{{\rm S}}(0)$ and reservoir vacuum state $|\{0_{k}\}\rangle\langle\{0_{k}\}|$. The task is now to calculate the two-time correlation function in the above equation.

The correlation function can be evaluated as
\begin{equation}
\langle\sigma_{x}(t)\sigma_{x}(t^{\prime})\rangle={\rm Tr}_{{\rm S}}\{\sigma_{x}{\rm Tr}_{{\rm R}}[U(t)U^{\dagger}(t^{\prime})\sigma_{x}\rho(t^{\prime})U(t^{\prime})U^{\dagger}(t)]\},
\end{equation}
where $\rho(t)$ is the
density matrix for the total system. This means that
the two-time correlation function can be obtained as an expectation
of $\sigma_{x}$ with respect to the reduced effective density operator,
$\Lambda_{{\rm S}}(t,t^{\prime})={\rm Tr}_{{\rm R}}[U(t)U^{\dagger}(t^{\prime})\sigma_{x}\rho(t^{\prime})U(t^{\prime})U^{\dagger}(t)]$.
Similarly to the reduced density matrix, the equation of motion for
$\Lambda_{{\rm S}}(t,t^{\prime})$ can be derived by using the Nakajima-Zwanzig
projection approach or a second-order perturbation calculation~\cite{mccutcheon,goan}. Given Hamiltonian (\ref{eq:Ham}) and zero temperature, we
derive the equations of motion for the effective density operator $\Lambda_{{\rm S}}(t,t^{\prime})$
and the reduced density operator $\rho_{{\rm S}}(t)={\rm Tr}_{{\rm R}}\rho(t)$ as follows:\begin{widetext}

\begin{eqnarray}
\frac{d}{dt}\Lambda_{{\rm S}}(t,t^{\prime}) & = & -i[H_{{\rm S}}(t),\Lambda_{{\rm S}}(t,t^{\prime})]-\int_{0}^{t-t^{\prime}}d\tau\{C(\tau)[\sigma_{x},\sigma_{x}(t,t-\tau)\Lambda_{{\rm S}}(t,t^{\prime})]-C^{\ast}(\tau)[\sigma_{x},\Lambda_{{\rm S}}(t,t^{\prime})\sigma_{x}(t,t-\tau)]\}\nonumber \\
 &  & -\int_{t-t^{\prime}}^{t}d\tau\{C(\tau)[\sigma_{x},\sigma_{x}(t,t^{\prime})\sigma_{x}(t,t-\tau)\rho_{{\rm S}}(t,t^{\prime})]-C^{\ast}(\tau)[\sigma_{x},\sigma_{x}(t,t^{\prime})\rho_{{\rm S}}(t,t^{\prime})\sigma_{x}(t,t-\tau)]\},\label{eq:correq}
\end{eqnarray}
\begin{eqnarray}
\frac{d}{dt}\rho_{{\rm S}}(t) & = & -i[H_{{\rm S}}(t),\rho_{{\rm S}}(t)]-\int_{0}^{t}d\tau\{C(\tau)[\sigma_{x},\sigma_{x}(t,t-\tau)\rho_{{\rm S}}(t)]+{\rm h.c.}\}.\label{eq:me}
\end{eqnarray}
\end{widetext}where
\begin{equation}
C(\tau)=\frac{1}{4}\int_{0}^{\infty}J(\omega)e^{-i\omega\tau}d\omega,
\end{equation}
\begin{equation}
\sigma_{x}(t,t^{\prime})=U_{{\rm S}}(t)U_{{\rm S}}^{\dagger}(t^{\prime})\sigma_{x}U_{{\rm S}}(t^{\prime})U_{{\rm S}}^{\dagger}(t),
\end{equation}
\begin{equation}
\rho_{{\rm S}}(t,t^{\prime})=U_{{\rm S}}(t)\rho_{{\rm S}}(t^{\prime})U_{{\rm S}}^{\dagger}(t).
\end{equation}
Here, $U_{{\rm S}}(t)={\cal T}_{\leftarrow}\exp\left[-i\int_{0}^{t}H_{{\rm S}}(\tau)d\tau\right]$
is the unitary evolution operator for the driven qubit only. The detailed
derivation of Eqs. (\ref{eq:correq}) and (\ref{eq:me}) are given
in Appendix \ref{sec:appme}. The present formalism of deriving equation of motion for the effective density operator is referred to as the nonMarkovian quantum regression theory~\cite{mccutcheon,goan}.

Clearly, the equations of motion for $\Lambda_{\rm S}(t,t^\prime)$ and $\rho_{\rm S}(t)$ are accurate up to the second order in the coupling strength between the qubit and the reservoir ($\lambda_k$). Thus, the master equation approach can be expected to give reliable results in sufficiently weak coupling regimes~\cite{lidar}. Furthermore, by
comparing Eq.~(\ref{eq:correq}) and (\ref{eq:me}), one finds that
the effective density operator and the density operator satisfy
different equations of motion. There exists an inhomogeneous term (the second line)  in Eq.~(\ref{eq:correq}). This is different from the usual Markovian case, in which the reduced density operator and effective density operator satisfy the same differential equations. To numerically calculate the spectrum, we rewrite the equations of motion in the Floquet picture, which is presented in appendix~\ref{sec:appFP}, and solve them with the Runge-Kutta method. On solving Eqs.~(\ref{eq:correq}) and (\ref{eq:me}), we can perform numerical integration for Eq.~(\ref{eq:sxsx}) and obtain the spectrum.

One notes that the photon number is determined by the correlation function $\langle\sigma_x(t)\sigma_x(t^\prime)\rangle$ instead of the correlation function $\langle \sigma_{+}(t)\sigma_{-}(t^\prime)\rangle$ with $\sigma_{\pm}=(\sigma_{x}\pm i\sigma_{y})/2$, which is widely used in the studies of the steady-state or time-dependent fluorescence spectra~\cite{scully,mollow,Renaud,Eberly}. This is because the RWA is not invoked when deriving $N(\omega_k,t)$, i.e., Eq.~(\ref{eq:sxsx}). If the qubit-reservoir coupling is assumed to take the RWA form, one finds that the photon number of the reservoir is related to $\langle \sigma_{+}(t)\sigma_{-}(t^\prime)\rangle$~\cite{corrrwa}.

\vspace{0.2cm}

\subsubsection{Photon number evaluated with the RWA}

Using the RWA, i.e., with the counter-rotating coupling $\frac{1}{2}\sum_{k}\lambda_{k}(b_{k}\sigma_{-}+b_{k}^{\dagger}\sigma_{+})$ omitted in the original Hamiltonian, one finds that~\cite{corrrwa}
\begin{equation}
N(\omega_{k},t)=\frac{\lambda_{k}^{2}}{4}\int_{0}^{t}\int_{0}^{t}\langle\sigma_{+}(t_{1})\sigma_{-}(t_{2})\rangle e^{-i\omega_{k}(t_{1}-t_{2})}dt_{1}dt_{2}, \label{eq:spsn}
\end{equation}
where $\sigma_{\pm}(t)={\cal U}^\dagger(t)\sigma_{\pm}{\cal U}(t)$ are the operators in the Heisenberg picture with ${\cal U}(t)$ being the evolution operator for the RWA Hamiltonian.

Using the same Nakajima-Zwanzig projection approach~\cite{mccutcheon}, the equations
of motion for the reduced effective density matrix $\Xi_{{\rm S}}(t,t^{\prime})={\rm Tr}_{{\rm R}}[{\cal U}(t){\cal U}^{\dagger}(t^{\prime})\sigma_{-}\rho(t^{\prime}){\cal U}(t^{\prime}){\cal U}^{\dagger}(t)]$
and the reduced density matrix $\rho_{{\rm S}}(t)$ can be given as
follows: \begin{widetext}
\begin{eqnarray}
\frac{d}{dt}\Xi_{{\rm S}}(t,t^{\prime}) & = & -i[H_{{\rm S}}(t),\Xi_{{\rm S}}(t,t^{\prime})]-\int_{0}^{t-t^{\prime}}d\tau\{C(\tau)[\sigma_{+},\sigma_{-}(t,t-\tau)\Xi_{{\rm S}}(t,t^{\prime})]-C^{\ast}(\tau)[\sigma_{-},\Xi_{{\rm S}}(t,t^{\prime})\sigma_{+}(t,t-\tau)]\}\nonumber \\
 &  & -\int_{t-t^{\prime}}^{t}d\tau\{C(\tau)[\sigma_{+},\sigma_{-}(t,t^{\prime})\sigma_{-}(t,t-\tau)\rho_{{\rm S}}(t,t^{\prime})]-C^{\ast}(\tau)[\sigma_{-},\sigma_{-}(t,t^{\prime})\rho_{{\rm S}}(t,t^{\prime})\sigma_{+}(t,t-\tau)]\},\label{eq:correqrwa}
\end{eqnarray}
\begin{equation}
\frac{d}{dt}\rho_{{\rm S}}(t)=-i[H_{{\rm S}}(t),\rho_{{\rm S}}(t)]-\int_{0}^{t}d\tau\{C(\tau)[\sigma_{+},\sigma_{-}(t,t-\tau)\rho_{{\rm S}}(t)]+{\rm h.c.}\},\label{eq:merwa}
\end{equation}
\end{widetext}where $\sigma_{\pm}(t,t^{\prime})=U_{{\rm S}}(t)U_{{\rm S}}^{\dagger}(t^{\prime})\sigma_{\pm}U_{{\rm S}}(t^{\prime})U_{{\rm S}}^{\dagger}(t).$
In a similar way, we solve Eqs.~(\ref{eq:correqrwa}) and (\ref{eq:merwa}) numerically and their numerical solutions are used to numerically evaluate the double integral in Eq.~(\ref{eq:spsn}).

It remains unclear whether the spectra obtained from the nonRWA and RWA treatments have discrepancy under certain conditions. This will be discussed in Sec.~\ref{sec:result}.

\section{Qubit dynamics and the validity of variational approach}\label{sec:dyn}
\begin{figure*}
  \includegraphics[width=2\columnwidth]{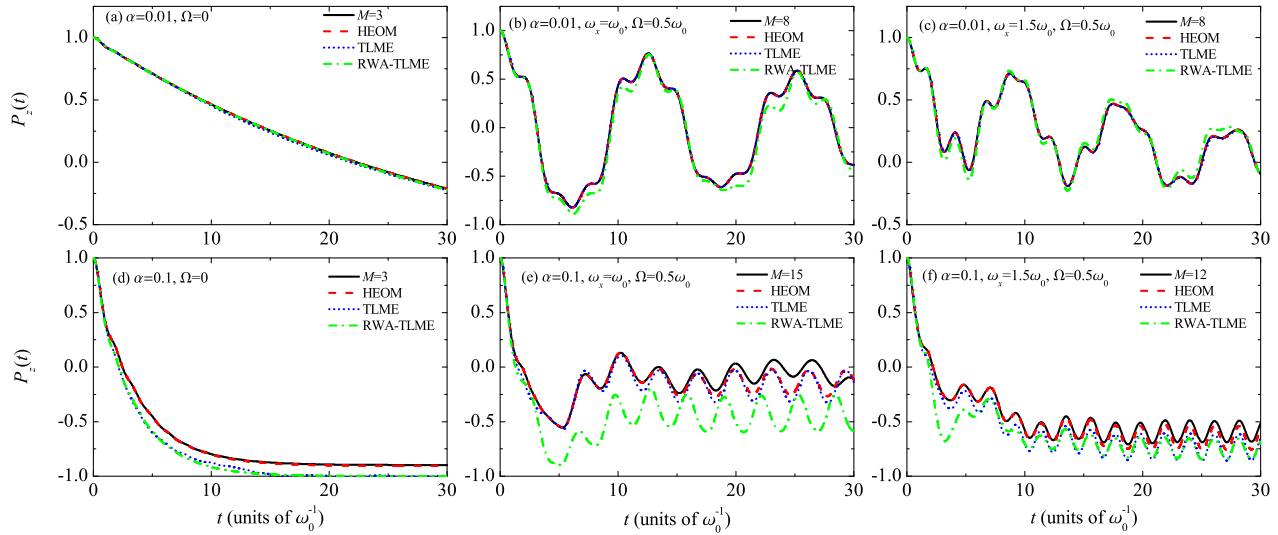}
  \caption{Dynamics of the population difference of qubit calculated from the variational approach, HEOM, TLME, and RWA-TLME for two values of $\alpha$. The solid line represents the multi-D$_1$ results with $N_b=150$ and $M$ being specified in the legend.}\label{fig1}
\end{figure*}

\begin{figure}
  \includegraphics[width=\columnwidth]{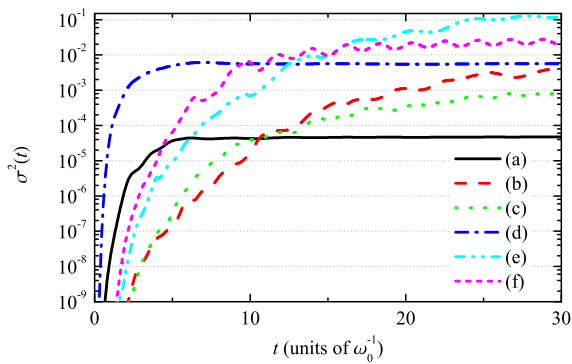}
  \caption{Time evolution of the deviation of the multi-D$_1$ results in Fig.~\ref{fig1}.}\label{fig2}
\end{figure}
In this section, we discuss the qubit dynamics and validity of the variational approach. To begin with, we examine the convergence of the variational solutions obtained from Eqs.~(\ref{eq:Aeq})-(\ref{eq:Geq}). The convergence is verified by adjusting $M$ and $N_b$. We first check the convergence with respect to the multiplicity $M$ by fixing $N_b$ and increasing $M$ until the increase in $M$ leads to negligible change in the qubit dynamics. We then check the convergence with respect to the number of modes $N_b$ by fixing $M$ and varying $N_b$ until the increase in $N_b$ causes negligible change in the qubit dynamics.
Referred to as the multi-D$_1$ results, the variational results presented in this work have been thoroughly scrutinized for convergence. Master Eqs.~(\ref{eq:me}) and (\ref{eq:merwa}) are referred to as the time local master equation (TLME) and RWA-TLME, respectively.
Next, we illustrate the performance of the multi-D$_1$ results. To this end, we make comparisons between the variational results with those from the master equations and the hierarchy equations of motion (HEOM)~\cite{PI3}. The details of HEOM can be found in Appendix~\ref{sec:heom}.

The HEOM results are first taken as a benchmark to validate the multi-D$_1$ results.
In Fig.~\ref{fig1}, we display the time-dependent population difference of the qubit for two values of $\alpha$. For $\alpha=0.01$, one finds that the multi-D$_1$ results are consistent with those from both HEOM and TLME. For $\alpha=0.1$, as shown in Figs.~\ref{fig1}(d) and \ref{fig1}(f), the multi-D$_1$ results have an acceptable accuracy as compared to the HEOM results and are more accurate than the TLME results. Figure~\ref{fig1}(e) shows that the mutli-D$_1$ outcome coincides with that of HEOM if $t<15\omega_0^{-1}$ but deviates from the latter if $t>15\omega_0^{-1}$. Interestingly, the TLME result has a satisfactory accuracy in the presence of the resonant strong driving even if $\alpha=0.1$. Nevertheless, the multi-D$_1$ ansatz has better performance than the master equations at short times.

To further illustrate the performance of the multi-D$_1$ ansatz, we proceed to calculate the ansatz deviation defined by:
\begin{equation}\label{eq:dev}
  \sigma^2(t)=\langle\delta(t)|\delta(t)\rangle/\omega_0^2,
\end{equation}
where
\begin{equation}
|\delta(t)\rangle=[i\partial_{t}-H(t)]|D_{M}(t)\rangle
\end{equation}
is the deviation vector.
The ansatz deviation $\sigma^2(t)$ measures how faithfully the trial state follows the Schr\"{o}dinger equation. Generally, the smaller the deviation $\sigma^2(t)$, the more accurate the variational solutions are, and $\sigma^2(t)=0$ if and only if the trial state is an exact solution to the Schr\"{o}dinger equation. An explicit expression for $\sigma^2(t)$ is derived in Appendix~\ref{sec:tdvp}. In Fig.~\ref{fig2}, we show $\sigma^2(t)$ of the multi-D$_1$ results in Fig.~\ref{fig1}.
We note that for vanishing driving, the ansatz deviation first increases and then approaches stable, acceptably small values, regardless of weak or strong coupling. However, in the presence of driving, the deviation keeps growing with time. This may lead to low accuracy of the multi-D$_1$ at long times. In addition, we have checked that for other driving parameters, e.g., $\{\omega_x=\omega_0,\Omega=1.5\omega_0\}$ and $\{\omega_0=0.56\omega_0,\Omega=\omega_0\}$, the deviation of the multi-D$_1$ results is similar as curve (e) in Fig.~\ref{fig2} when $\alpha=0.1$. However, the deviation is found to be acceptably small [$\sigma^2(t)<10^{-2}$] for $\alpha=0.1$ at $t=30\omega_0^{-1}$ (the qubit already relaxes to its steady state) if the driving is relatively weak, e.g., $\Omega\sim0.1\omega_0$.

Some remarks are due on the validity of the variational approach based on the mutli-D$_1$ trial state as well as that of the master equations. First of all, our time-dependent variation is particularly accurate for short time dynamics in all parameter regimes. Secondly, the variational approach yields numerical convergence to a more accurate steady state than the master equation in the presence of vanishing driving, relatively weak, and far-off-resonant driving, and in the strong-coupling regime, while in the presence of resonant strong driving, such convergence is elusive in the variational approach. Thirdly, even in the strong-coupling regime, the TLME may have a relatively good performance under the resonant strong driving condition but is inaccurate in the presence of vanishing driving, relatively weak or far-off-resonant driving. The RWA-TLME is generally inconsistent with the TLME in the presence of the strong driving.

Before ending this section, we would like to discuss dynamical features in the moderately weak and strong coupling regime.
Figures~\ref{fig1}(a)-\ref{fig1}(c) show that the qubit spontaneously decays in the moderately weak coupling regime and without driving,
while exhibiting damped Rabi oscillations in the presence of driving.
Figure~\ref{fig1}(d) shows that in the strong-coupling regime and without driving, the qubit spontaneously decays into a steady state after a relatively short period of time. The qubit is not found in its bare ground state at long times, revealing that the qubit is dressed by photons in the strong-coupling regime~\cite{diaz,zhengepjb}. It is seen that this effect cannot be captured by the second-order master equation. Figures~\ref{fig1}(e)-\ref{fig1}(f) show that with driving, the qubit exhibits oscillatory behavior at variance with the Rabi oscillation in the presence of strong qubit-reservoir coupling despite that $\Omega=0.5\omega_0$. Those results imply that the Rabi oscillation of qubit can be modified significantly by a strongly dissipative reservoir. Therefore, in the strong-coupling regime, we may expect that the fluorescence spectrum deviates substantially from the typical Mollow triplet.

\section{Photon number dynamics and time-dependent fluorescence spectrum}\label{sec:result}

In this section, using the three aforementioned approaches, we calculate the time-dependent photon number and the fluorescence spectrum, in an effort to probe the reservoir dynamics. Computation of photon number dynamics based on Eqs.~(\ref{eq:sxsx}), (\ref{eq:correq}), and (\ref{eq:me}) is referred to as the TLME approach, and calculation based on Eqs.~(\ref{eq:spsn}), (\ref{eq:correqrwa}), and (\ref{eq:merwa}) is referred to as the RWA-TLME approach. Discrepancy among the three approaches, together with computational consistency and robustness of the results obtained, will be addressed first, which is followed by discussion of time-dependent spectra.

\subsection{Comparison between variational and master-equation approaches}

\begin{figure*}
  \includegraphics[width=2\columnwidth]{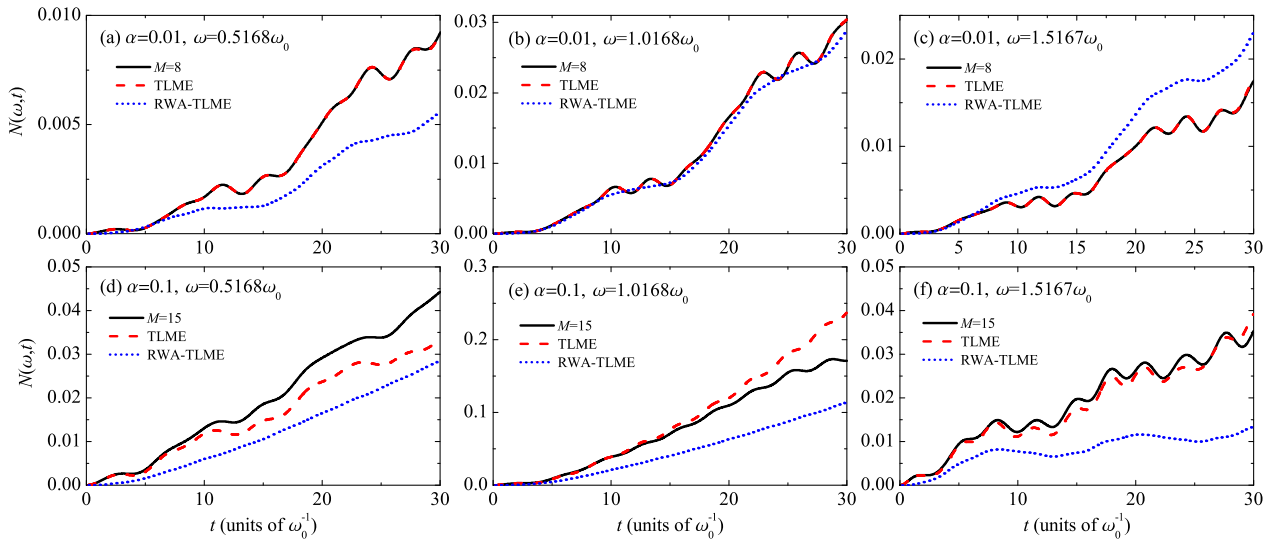}
  \caption{Time evolution of photon numbers at three discrete modes for two values of $\alpha$. The parameters are: $\omega_x=\omega_0$, $\Omega=0.5\omega_0$, and $\omega_0=0.2\omega_c$. The qubit is initially in the ground state. The solid lines are the multi-D$_1$ results with $N_b=150$ and $M$ being specified in the legends. }\label{fig3}
\end{figure*}

\begin{figure*}
  \includegraphics[width=2\columnwidth]{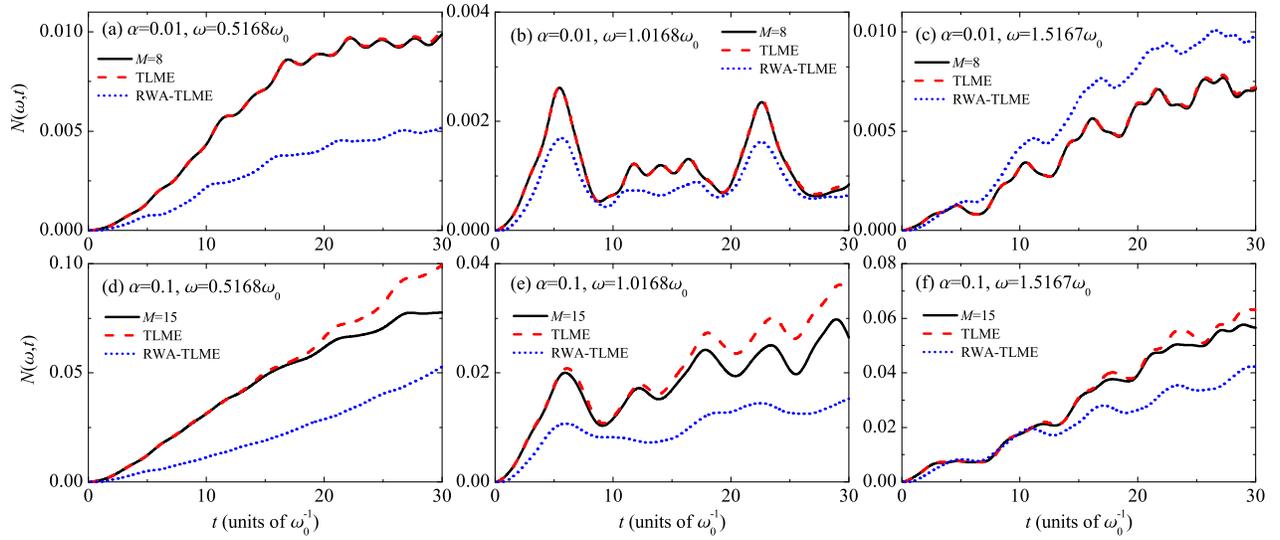}
  \caption{Time evolution of photon numbers at three discrete modes for two values of $\alpha$. The parameters are: $\omega_x=0.56\omega_0$, $\Omega=\omega_0$, and $\omega_0=0.2\omega_c$. The qubit is initially in the ground state. The solid lines are the multi-D$_1$ results with $N_b=150$ and $M$ being specified in the legends.}\label{fig4}
\end{figure*}

In this section, we examine the discrepancy between the variational and master-equation approaches. Without loss of generality, we consider the time evolution of photon numbers in three discrete modes with the frequencies $\omega=0.5168\omega_0$, $1.0168\omega_0$, and $1.5167\omega_0$, corresponding to the 16th, 31th, and 46th bosonic modes obtained in a linear discretization of the reservoir spectral density with a total number of $N_b=150$ modes. Figure~\ref{fig3} displays the time-dependent photon number $N(\omega,t)$ with the three chosen frequencies $\omega$ for the case of $\omega_x=\omega_0$, $\Omega=0.5\omega_0$, and two values of $\alpha$. The qubit is initially in its ground state. We first concentrate on the comparison between the TLME and multi-D$_1$ results. Figures~\ref{fig3}(a)-\ref{fig3}(c) show that when $\alpha=0.01$, the multi-D$_1$ curves perfectly coincide with the TLME curves. Figures~\ref{fig3}(d)-\ref{fig3}(f) show that when $\alpha=0.1$, perfect agreement between the multi-D$_1$ and TLME results is only found at short times.
Discrepancy between the multi-D$_1$ and the TLME results become apparent at long times, which is attributed to the inaccuracy of the multi-D$_1$ ansatz. Nevertheless, the two are qualitatively consistent with each other. In Fig.~\ref{fig4} we show $N(\omega,t)$ as a function of $t$ for a lower driving frequency and a stronger driving strength while keeping the other parameters the same as Fig.~\ref{fig3}. When comparing the multi-D$_1$ and TLME results, one finds a similar situation where the discrepancy between them is vanishingly small for $\alpha=0.01$ and is of considerable magnitude for $\alpha=0.1$.

\begin{figure*}
  \includegraphics[width=2\columnwidth]{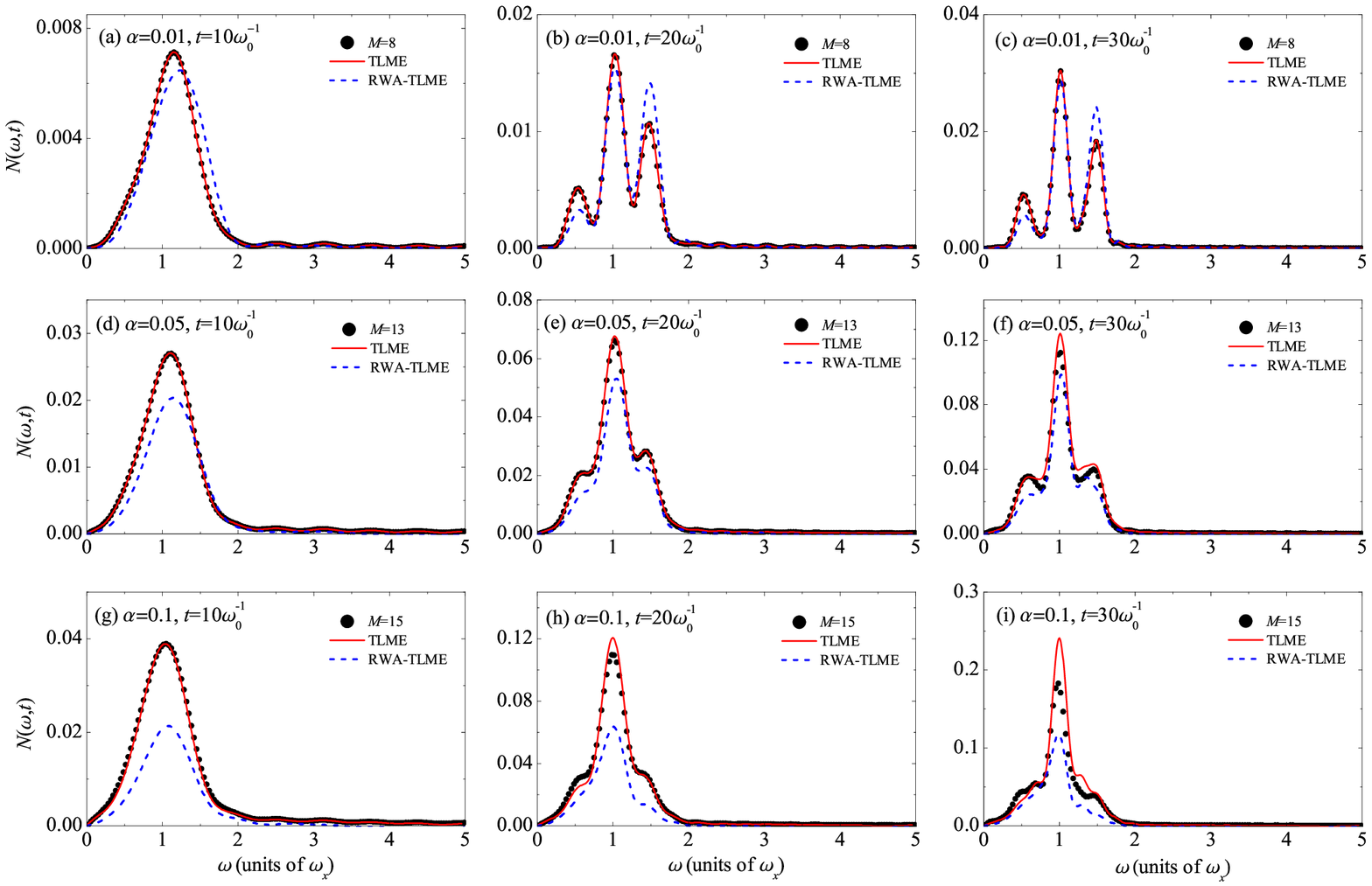}
  \caption{Time-dependent fluorescence spectrum as a function of $\omega$ at fixed times  for $\omega_x=\omega_0$, $\Omega=0.5\omega_0$, $\omega_0=0.2\omega_c$, and three values of $\alpha$. The qubit is initially in the ground state. The scatters represent the multi-D$_1$ results with $N_b=150$ and $M$ specified in the legends.}\label{fig5}
\end{figure*}

\begin{figure*}
 \includegraphics[width=2\columnwidth]{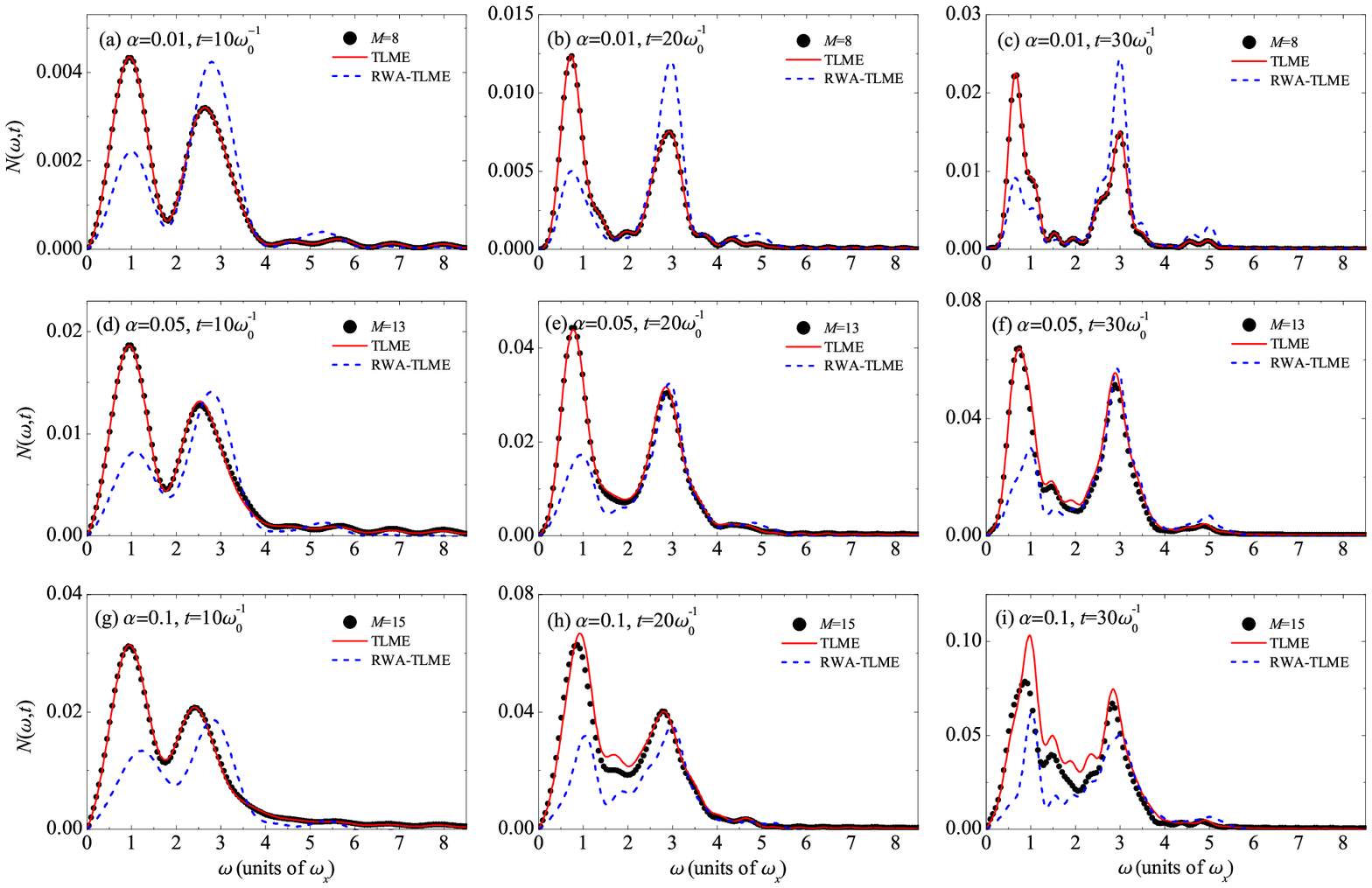}
  \caption{Time-dependent fluorescence spectrum as a function of $\omega$ at fixed times for $\omega_x=0.56\omega_0$, $\Omega=\omega_0$, $\omega_0=0.2\omega_c$, and three values of $\alpha$. The qubit is initially in the ground state. The scatters represent the multi-D$_1$ results with $N_b=150$ and $M$ specified in the legends.}\label{fig6}
\end{figure*}

\begin{figure*}
 \includegraphics[width=2\columnwidth]{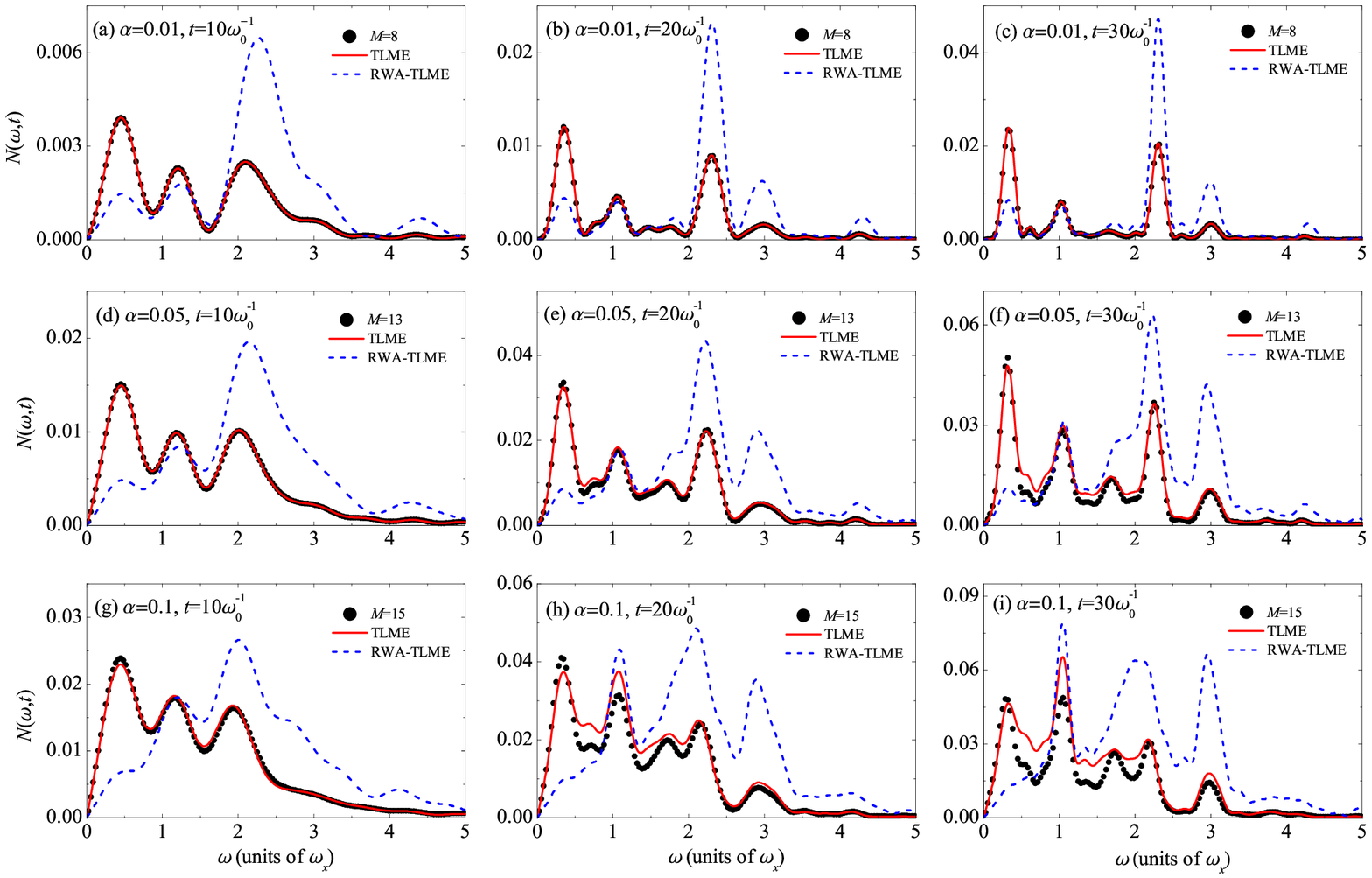}
  \caption{Time-dependent fluorescence spectrum as a function of $\omega$ at fixed times for $\omega_0=0.2\omega_c$, $\omega_x=\omega_0$, $\Omega=1.5\omega_0$, and three values of $\alpha$. The qubit is initially in the ground state. The scatters represent the multi-D$_1$ results with $N_b=150$ and $M$ specified in the legends.}\label{fig7}
\end{figure*}

Apart from the three discrete frequencies considered, it is easy to verify whether the master equation and variational approaches are consistent for other discrete modes. To this end, we plot $N(\omega,t)$ as a function of the reservoir frequency $\omega$ at given times, i.e., the time-dependent fluorescence spectrum. The results from three methods are shown in Figs.~\ref{fig5}-\ref{fig7}. In each figure, the coupling strength $\alpha$ ranges from $0.01$ to $0.1$. The top panels in Figs.~\ref{fig5}-\ref{fig7} show that for $\alpha=0.01$ the multi-D$_1$ results are in excellent agreement with the TLME results, while at $\alpha=0.05$ or $\alpha=0.1$, the TLME and multi-D$_1$ results agree well with each other at $t=10\omega_0^{-1}$, but the significant discrepancy between them appears at $t=30\omega_0^{-1}$, as demonstrated in the middle and bottom panels in Figs.~\ref{fig5}-\ref{fig7}. Nevertheless, there is no essential difference between the profiles of the multi-D$_1$ and the TLME spectra.

We move to compare the RWA-TLME results with the TLME and multi-D$_1$ results. It is evident that the RWA-TLME results are different from their non-RWA counterparts. First, Figs.~\ref{fig3} and~\ref{fig4} show that the RWA-TLME curves are smoother than the non-RWA curves, namely, the fast oscillatory behavior of photon number is not captured by the RWA. Second, more importantly, there is a considerable discrepancy between the RWA and non-RWA photon numbers at certain frequencies. For example, if $\alpha=0.01$, one finds the significant discrepancy at $\omega=0.5168\omega_0$ and $\omega=1.5167\omega_0$ but very little disagreement at $\omega=1.0168\omega_0$,
suggesting that in the weak coupling regime, RWA may be a bad approximation at the frequencies far from the qubit frequency $\omega_0$ but may still yield reasonably accurate phonon number dynamics at the frequencies close to the qubit frequency.
For $\alpha=0.1$, the RWA and non-RWA results have significant discrepancies as shown in the lower panels in Figs.~\ref{fig3} and ~\ref{fig4}. Although both TLME and RWA-TLME results are somewhat inaccurate at long times because they are second-order perturbations, there is a growing discrepancy between the RWA and non-RWA results as $\alpha$ increases.

Figs.~\ref{fig5}-\ref{fig7} compare the RWA-TLME and nonRWA results by plotting the photon number $N(\omega,t)$ as a function of $\omega$ at given time $t$. One finds that from the weak to strong coupling regime, the RWA-TLME approach predicts a transient photon number distribution over a  frequency different from those from the other two methods. The discrepancy between the RWA-TLME and TLME results can be attributed to the effect of the counter-rotating coupling between the qubit and reservoir, which is not taken into account in the RWA-TLME results. These findings suggest that under the strong driving condition the counter-rotating qubit-bath coupling has non-negligible contributions to the photon number dynamics even in the weak-coupling regime.

One may ask whether the RWA-TLME and TLME results become indistinguishable under certain conditions. To answer this question, we have used the RWA-TLME and TLME approaches to calculate $N(\omega,t)$ for $\omega_x=\omega_0$ and various values of $\alpha$ ranging from $10^{-3}$ to $10^{-5}$. We find that as long as $\Omega$ is comparable with $\omega_0$, the RWA-TLME and TLME results are different (the discrepancy is similar as those shown in Figs.~\ref{fig3} and \ref{fig4} and thus they are not presented), namely, these two results are different in the regime of $\alpha\ll1$ and $\Omega/\omega_0\sim1$. However, when $\Omega/\omega_0\ll1$, the RWA-TLME and TLME results become indistinguishable provided that $\alpha\ll1$, i.e., the effect of the counter-rotating coupling is negligible in the weak-coupling and the weak-driving limit. This is the often considered regime where both the spontaneous decay rate and Rabi frequency are far smaller than the transition frequency.

\subsection{Spectral features}

In this section, we focus on the features of the time-dependent fluorescence spectra and seek to understand them
in terms of the photon number dynamics.
Figures~\ref{fig5}-\ref{fig7} show that regardless of the RWA employment, the time-dependent fluorescence spectrum is generally asymmetric as we go from the weak to strong coupling regime. In general, we find that it is nontrivial to obtain symmetric time-dependent spectrum with the present model when the driving is strong. This can be intuitively understood by considering the photon number dynamics. If the time-dependent spectrum is symmetric about a central frequency, the photon number dynamics should also have mirror symmetry about the central frequency. However, this situation cannot be trivially realized except for some limiting cases. It is clear that the coupling strength between the qubit and bosonic mode varies from mode to mode, as given by the spectral density $J(\omega)$. Consequently, one has no reason to expect that the photon number dynamics of one mode is exactly the same as that of another. In fact, the photon number dynamics varies from mode to mode as illustrated in Figs.~\ref{fig3} and \ref{fig4}. This explanation for the asymmetry is different from the previous attempts based on the quantum optical master equation, which ascribes the asymmetry of the time-dependent spectrum to the ``turn-on'' effect of interaction between emitter and laser field~\cite{Renaud,Eberly}. In addition, we can state one limiting case where the symmetric time-dependent spectrum is observed. When $\Omega\ll\omega_0$ and $\alpha\ll1$, the equations of motion (\ref{eq:correqrwa}) and (\ref{eq:merwa}) can be well approximated by the quantum optical master equation. In such a situation, the symmetric transient fluorescence spectrum can be observed under certain conditions inferred from the previous works~\cite{Renaud,Eberly}.

Let us analyze how the spectrum varies with $\alpha$. Figures~\ref{fig5}-\ref{fig7} show that at a given time, the intensity of spectrum is much greater in the strong-coupling regime than in the weak coupling regime, which results from the fact that the increase of coupling strength $\alpha$ leads to the enhancement of  spontaneous decay rate. The last columns of
Figs.~\ref{fig5}-\ref{fig7} show that the emission bands are connected and form an integrated whole in the strong coupling regime; in contrast, the emission bands are well separated in the weak coupling regime.
For instance, Fig.~\ref{fig5} shows that the increase of $\alpha$ causes the spectrum to change from a Mollow-triplet-like structure into the structure that a single peak is integrated with a very broad band at long times.
This means that in the weak-coupling regime, few photons can be scattered into the modes at the gaps among the emission bands. However, in the strong-coupling regime, these modes can be considerably populated with photons. The present results suggest that the spectral profiles can be significantly modified by the strong dissipation.

\begin{figure*}
 \includegraphics[width=1\columnwidth]{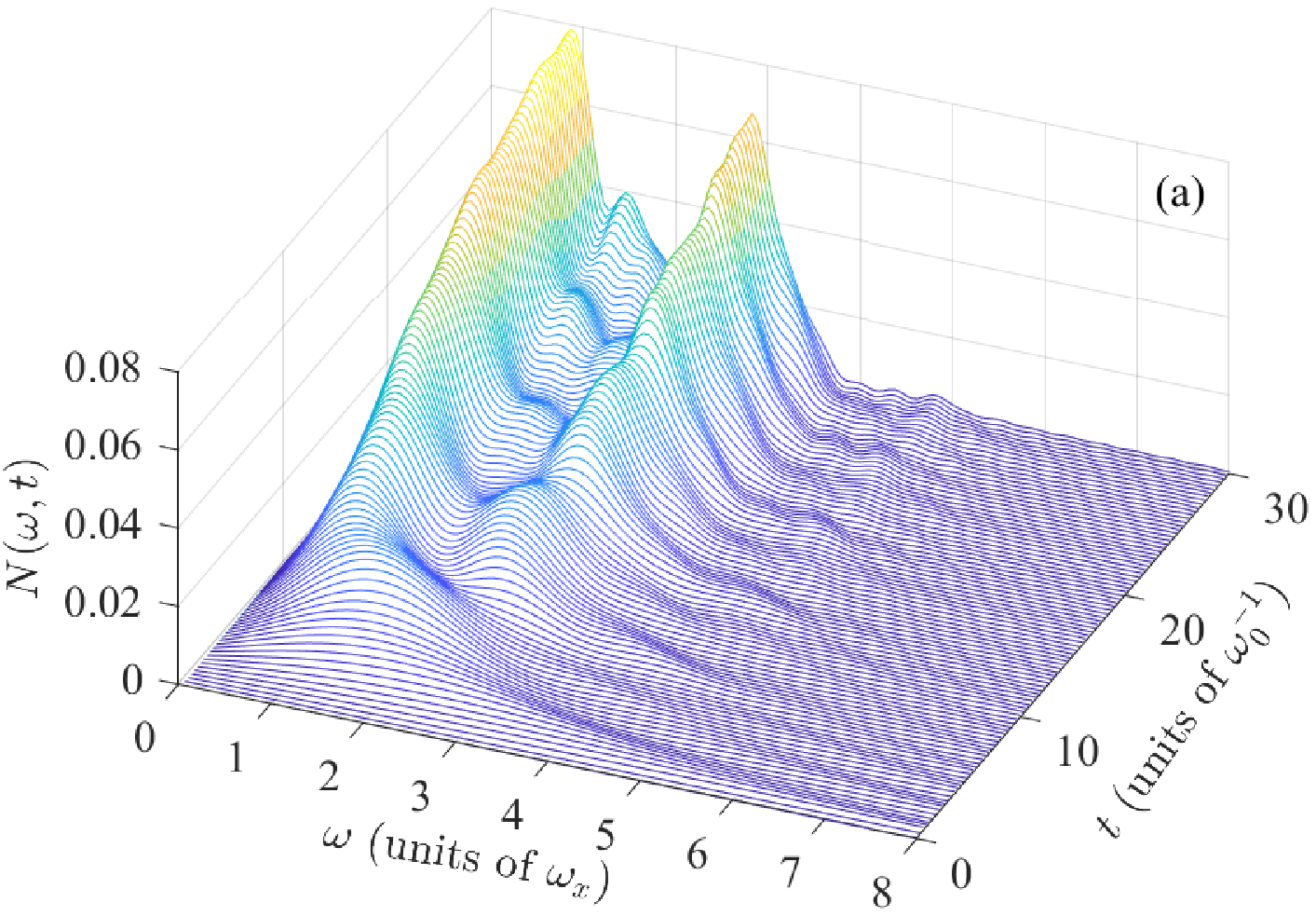}
 \includegraphics[width=1\columnwidth]{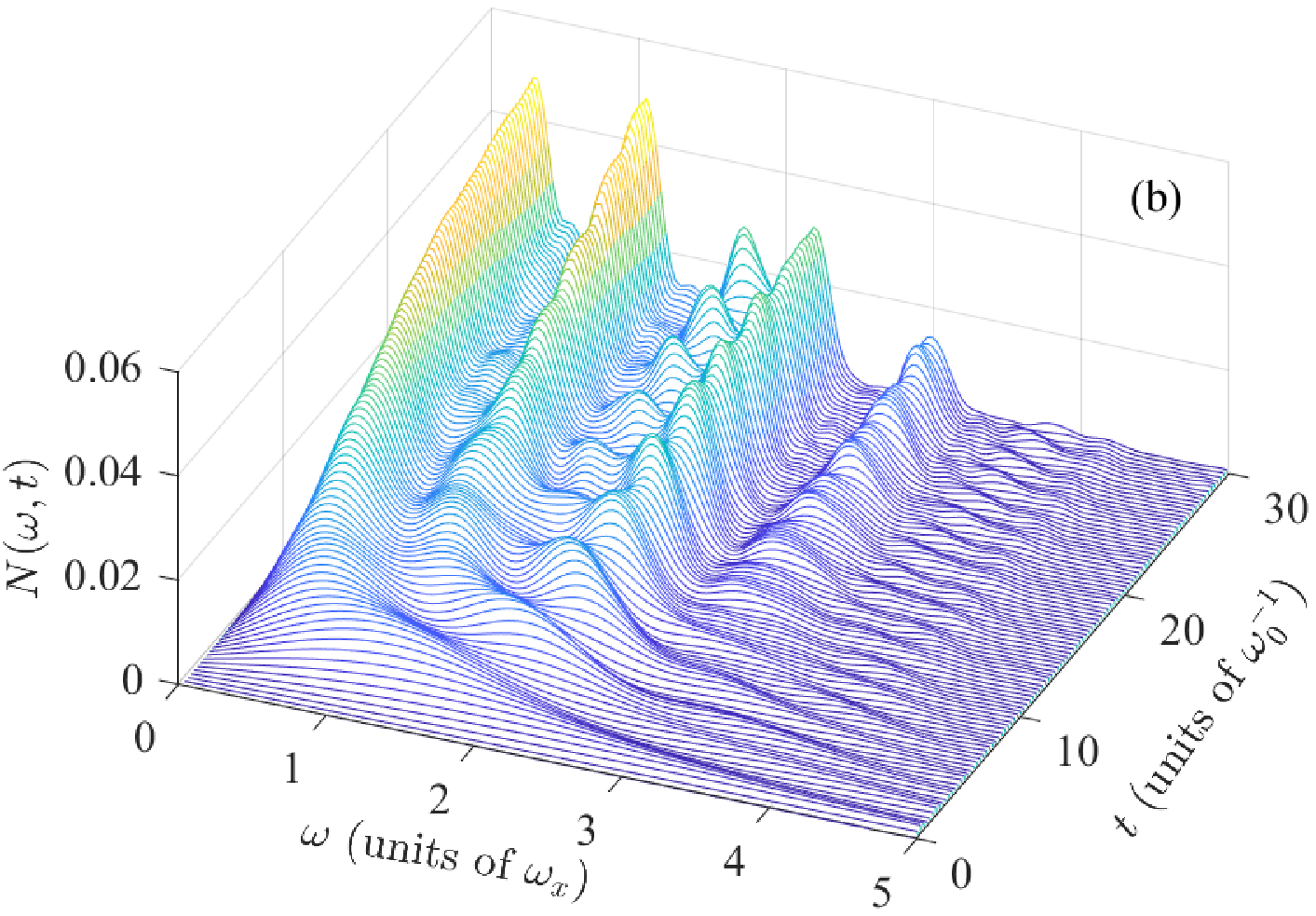}
 \caption{Time-dependent fluorescence spectrum versus frequency $\omega$ and time $t$ from the multi-D$_1$ results for $\alpha=0.1$. The parameters and initial condition used in (a) and (b) are the same as in Figs.~\ref{fig6} and~\ref{fig7}, respectively.}\label{fig8}
\end{figure*}

Our approach here also allows for studying the fluorescence spectrum in the strong-driving regime. Figs.~\ref{fig6} and \ref{fig7} show that the spectral profiles can be dramatically different from the usual Mollow triplet when the driving strength equals to or exceeds the transition frequency of the emitter. At long times, the spectrum is found to primarily consist of two components around $n\omega_x$ $(n=1,3)$, which originate in the single- and three-photon processes. It is seen that the two components have comparable intensities. This suggests that the three-photon process plays an equal role as the single-photon process under strong driving conditions, inferring that the multiphoton processes can substantially modify the spectral profiles when the driving is strong.
In Figs.~\ref{fig8}(a) and \ref{fig8}(b), we plot the fluorescence spectrum, obtained from our variational approach, as a function of frequency $\omega$ and time $t$ for $\alpha=0.1$ with other parameters being the same as in Figs.~\ref{fig6} and \ref{fig7}, respectively.
These plots reveal the oscillatory behavior for each mode and detailed spectral evolution. One notes that high-frequency modes generally oscillate faster than low-frequency modes. In addition, the modes with frequencies greater than $3\omega_x$ are slightly excited even under strong driving.

\section{Conclusions}\label{sec:con}
To summarize, we have studied time-dependent fluorescence spectrum in the moderately weak to strong coupling regime by using the Dirac-Frenkel variational principle and the multiple Davydov D$_1$ ansatz. The validity of this method is shown by the comparison of the reduced dynamics with that of the HEOM and by the calculation of the ansatz deviation. In contrast with the master-equation approach, this method allows us to bypass two-time correlation functions and directly evaluate the number of scattered photons. We have compared the variational results with those from the two versions of TLME approach: one is based on RWA, and the other is without RWA. In the moderately weak-coupling regime, the variational approach and the TLME approach are found to agree with each other. However, in the strong-coupling regime, results from the two methods diverge at long times. In the case of resonant strong driving, the discrepancy is caused by the inadequate accuracy of the multi-D$_1$ trial state. In the cases of vanishing, weak, and far-off-resonant driving, the discrepancy is attributed to the breakdown of the second-order perturbation used in the master equation approach.
By comparing the RWA-TLME results with those of TLME and multi-D$_1$,
we have illustrated that the counter-rotating coupling between qubit and reservoir has considerable contributions to photon number dynamics and spectra when driving is comparable with the transition frequency of qubit.
Employing the three methods, we have shown that time-dependent fluorescence spectra are generally asymmetric. This can be understood from the viewpoint of photon number dynamics. In addition, we have shown that the strong dissipation and/or multiphoton processes can cause the spectral profile to be substantially different from the Mollow triplet when the driving strength is comparable with to or exceeds the transition frequency.

Our variational formalism equipped with the multiple Davydov D$_1$ ansatz provides a flexible way to compute the time-dependent spectrum, which captures not only the qubit dynamics but also the field dynamics. The present formalism is capable of treating relatively complicated models of interest in quantum optics. For instance, we can extend the ansatz to study bosonic dynamics when the multiple emitters interact with a common bath~\cite{hzk-zfl}. We can also study the bosonic dynamics when two bosonic baths interact with an emitter. An concrete example of the latter is semiconductor quantum dots whose electromagnetic and phonon environments should be taken into account. In addition, although only harmonic driving is considered here, the formalism is applicable to studying pulse or aperiodic driving fields.

\begin{acknowledgments}
The authors thank Zhiguo L\"{u} and Lu Wang
for useful discussion. Support from the National Natural Science
Foundation of China (Grants No.~11647082 and No.~11774311), and the Singapore
Ministry of Education Academic Research Fund Tier 1 (Grant
No.~RG190/18) is gratefully acknowledged.
\end{acknowledgments}

\appendix

\section{Equations of motion for the variational parameters and norm of the deviation vector\label{sec:tdvp}}

The variation of $\langle D_{M}(t)|$, which is the adjoint state
of $|D_{M}(t)\rangle$, can be obtained as follows:

\begin{eqnarray}
\langle\delta D_{M}(t)| & = & \sum_{l=1}^{M}\bigg\{\delta A_{l}^{\ast}\langle+|\langle f_{l}|+A_{l}^{\ast}\langle+|\langle f_{l}|\sum_{p}b_{p}\delta f_{lp}^{\ast}\nonumber \\
 &  & +\delta B_{l}^{\ast}\langle-|\langle g_{l}|+B_{l}^{\ast}\langle-|\langle g_{l}|\sum_{p}b_{p}\delta g_{lp}^{\ast}\bigg\}.\nonumber\\
\end{eqnarray}
One readily derives the derivative of $|D_{M}(t)\rangle$ with respect
to $t$,
\begin{eqnarray}
|\dot{D}_{M}(t)\rangle & = & \sum_{n=1}^{M}\left\{\dot{A}_{n}|+\rangle|f_{n}\rangle+A_{n}|+\rangle\left[\sum_{k}\dot{f}_{nk}b_{k}^{\dagger}\right]|f_{n}\rangle\right.\nonumber \\
 &  & \left.+\dot{B}_{n}|-\rangle|g_{n}\rangle+B_{n}|-\rangle\left[\sum_{k}\dot{g}_{nk}b_{k}^{\dagger}\right]|g_{n}\rangle\right\}.\nonumber\\
\end{eqnarray}
The Dirac-Frenkel time-dependent variational principle leads to the
equations of motion:
\begin{equation}
\langle+|\langle f_{l}|i\partial_{t}|D_{M}(t)\rangle=\langle+|\langle f_{l}|H(t)|D_{M}(t)\rangle,
\end{equation}
\begin{equation}
\langle-|\langle g_{l}|i\partial_{t}|D_{M}(t)\rangle=\langle-|\langle g_{l}|H(t)|D_{M}(t)\rangle,
\end{equation}
\begin{equation}
\langle+|\langle f_{l}|b_{p}i\partial_{t}|D_{M}(t)\rangle=\langle+|\langle f_{l}|b_{p}H(t)|D_{M}(t)\rangle,
\end{equation}
\begin{equation}
\langle-|\langle g_{l}|b_{p}i\partial_{t}|D_{M}(t)\rangle=\langle-|\langle g_{l}|b_{p}H(t)|D_{M}(t)\rangle.
\end{equation}
By substituting the explicit forms of $|D_{M}(t)\rangle$ and $H(t)$ into above equations, one simply derives the equations of motion in the main text.

To quantify how faithfully the variational results follow the Schr\"{o}dinger
equation, we calculate the deviation defined in Eq.~(\ref{eq:dev}), which is fully determined by the inner product of the deviation vector and can be derived straightforwardly
as follows:
\begin{eqnarray}
\langle\delta(t)|\delta(t)\rangle & = & \left\langle \dot{D}_{M}(t)|\dot{D}_{M}(t)\right\rangle +\left\langle D_{M}(t)|H(t)^{2}|D_{M}(t)\right\rangle \nonumber\\
 &  & -2{\rm Im}\langle\dot{D}_{M}(t)|H(t)|D_{M}(t)\rangle,
\end{eqnarray}
where
\begin{widetext}
\begin{eqnarray}
\left\langle \dot{D}_{M}(t)|\dot{D}_{M}(t)\right\rangle  & = & \sum_{l,n}\left[\left(\dot{A}_{l}^{\ast}\dot{A}_{n}+\dot{A}_{l}^{\ast}A_{n}\sum_{k}f_{lk}^{\ast}\dot{f}_{nk}+A_{l}^{\ast}\dot{A}_{n}\sum_{k}\dot{f}_{lk}^{\ast}f_{nk}\right)S_{ln}^{(f,f)}+\left(\dot{B}_{l}^{\ast}\dot{B}_{n}+\dot{B}_{l}^{\ast}B_{n}\sum_{k}g_{lk}^{\ast}\dot{g}_{nk}\right.\right.\nonumber \\
 &  & \left.+B_{l}^{\ast}\dot{B}_{n}\sum_{k}\dot{g}_{lk}^{\ast}g_{nk}\right)S_{ln}^{(g,g)}+A_{l}^{\ast}A_{n}\left(\sum_{k}\dot{f}_{lk}^{\ast}\dot{f}_{nk}+\sum_{k,q}f_{lk}^{\ast}\dot{f}_{nk}\dot{f}_{lq}^{\ast}f_{nq}\right)S_{ln}^{(f,f)}\nonumber \\
 &  & \left.+B_{l}^{\ast}B_{n}\left(\sum_{k}\dot{g}_{lk}^{\ast}\dot{g}_{nk}+\sum_{k,q}g_{lk}^{\ast}\dot{g}_{nk}\dot{g}_{lq}^{\ast}g_{nq}\right)S_{ln}^{(g,g)}\right],
\end{eqnarray}
\begin{eqnarray}
\langle\dot{D}_{M}(t)|H(t)|D_{M}(t)\rangle & = & \sum_{l,n}\left\{ \frac{\omega_{0}}{2}\dot{A}_{l}^{\ast}B_{n}S_{ln}^{(f,g)}+\dot{A}_{l}^{\ast}A_{n}\left[\Omega\cos(\omega_{x}t)+\sum_{k}\omega_{k}f_{lk}^{\ast}f_{nk}+\sum_{k}\frac{\lambda_{k}}{2}(f_{lk}^{\ast}+f_{nk})\right]S_{ln}^{(f,f)}\right.\nonumber \\
 &  & +\frac{\omega_{0}}{2}A_{l}^{\ast}B_{n}\sum_{k}\dot{f}_{lk}^{\ast}g_{nk}S_{ln}^{(f,g)}+A_{l}^{\ast}A_{n}\left[\Omega\cos(\omega_{x}t)\sum_{k}\dot{f}_{lk}^{\ast}f_{nk}+\sum_{k}\omega_{k}\dot{f}_{lk}^{\ast}f_{nk}\right.\nonumber \\
 &  & \left.+\sum_{k,q}\omega_{k}\dot{f}_{lq}^{\ast}f_{nq}f_{lk}^{\ast}f_{nk}+\sum_{k}\frac{\lambda_{k}}{2}\dot{f}_{lk}^{\ast}+\sum_{k,q}\frac{\lambda_{k}}{2}(f_{lk}^{\ast}+f_{nk})\dot{f}_{lq}^{\ast}f_{nq}\right]S_{ln}^{(f,f)}\nonumber \\
 &  & +\frac{\omega_{0}}{2}\dot{B}_{l}^{\ast}A_{n}S_{ln}^{(g,f)}-\dot{B}_{l}^{\ast}B_{n}\left[\Omega\cos(\omega_{x}t)-\sum_{k}\omega_{k}g_{lk}^{\ast}g_{nk}+\sum_{k}\frac{\lambda_{k}}{2}(g_{lk}^{\ast}+g_{nk})\right]S_{ln}^{(g,g)}\nonumber \\
 &  & +\frac{\omega_{0}}{2}B_{l}^{\ast}A_{n}\sum_{k}\dot{g}^{\ast}_{lk}f_{nk}S_{ln}^{(g,f)}-B_{l}^{\ast}B_{n}\left[\Omega\cos(\omega_{x}t)\sum_{k}\dot{g}_{lk}^{\ast}g_{nk}-\sum_{k}\omega_{k}\dot{g}_{lk}^{\ast}g_{nk}\right.\nonumber \\
 &  & \left.\left.-\sum_{k,q}\omega_{k}g_{lk}^{\ast}g_{nk}\dot{g}_{lq}^{\ast}g_{nq}+\sum_{k}\frac{\lambda_{k}}{2}\dot{g}_{lk}^{\ast}+\sum_{k,q}\frac{\lambda_{k}}{2}(g_{lk}^{\ast}+g_{nk})\dot{g}_{lq}^{\ast}g_{nq}\right]S_{ln}^{(g,g)}\right\} ,
\end{eqnarray}
\begin{eqnarray}
\left\langle D_{M}(t)|H(t)^{2}|D_{M}(t)\right\rangle  & = & \sum_{l,n}\left\{ \left[\frac{1}{4}\omega_{0}^{2}+\Omega^{2}\cos^{2}(\omega_{x}t)\right](A_{l}^{\ast}A_{n}S_{ln}^{(f,f)}+B_{l}^{\ast}B_{n}S_{ln}^{(g,g)})+\omega_{0}A_{l}^{\ast}B_{n}\sum_{k}\omega_{k}f_{lk}^{\ast}g_{nk}S_{ln}^{(f,g)}\right.\nonumber \\
 &  & +\omega_{0}B_{l}^{\ast}A_{n}\sum_{k}\omega_{k}g_{lk}^{\ast}f_{nk}S_{ln}^{(g,f)}+2\Omega\cos(\omega_{x}t)\left[A_{l}^{\ast}A_{n}\sum_{k}\left(\omega_{k}f_{lk}^{\ast}f_{nk}+\frac{\lambda_{k}}{2}(f_{lk}^{\ast}+f_{nk})\right)S_{ln}^{(f,f)}\right.\nonumber \\
 &  & \left.-B_{l}^{\ast}B_{n}\sum_{k}\left(\omega_{k}g_{lk}^{\ast}g_{nk}-\frac{\lambda_{k}}{2}(g_{lk}^{\ast}+g_{nk})\right)S_{ln}^{(g,g)}\right]\nonumber \\
 &  & +A_{l}^{\ast}A_{n}\left(\sum_{k}\omega_{k}^{2}f_{lk}^{\ast}f_{nk}+\sum_{k}\frac{\lambda_{k}^{2}}{4}+\sum_{k,q}\omega_{k}\omega_{q}f_{lk}^{\ast}f_{nk}f_{lq}^{\ast}f_{nq}+\sum_{k,q}\frac{\lambda_{k}\lambda_{q}}{4}(f_{lk}^{\ast}+f_{nk})(f_{lq}^{\ast}+f_{nq})\right.\nonumber \\
 &  & \left.+\sum_{k}\frac{\omega_{k}\lambda_{k}}{2}(f_{lk}^{\ast}+f_{nk})+\sum_{k,q}\omega_{k}\lambda_{q}f_{lk}^{\ast}f_{nk}(f_{lq}^{\ast}+f_{nq})\right)S_{ln}^{(f,f)}\nonumber \\
 &  & +B_{l}^{\ast}B_{n}\left(\sum_{k}\omega_{k}^{2}g_{lk}^{\ast}g_{nk}+\sum_{k}\frac{\lambda_{k}^{2}}{4}+\sum_{k,q}\omega_{k}\omega_{q}g_{lk}^{\ast}g_{nk}g_{lq}^{\ast}g_{nq}+\sum_{k,q}\frac{\lambda_{k}\lambda_{q}}{4}(g_{lk}^{\ast}+g_{nk})(g_{lq}^{\ast}+g_{nq})\right.\nonumber \\
 &  & \left.\left.-\sum_{k}\frac{\omega_{k}\lambda_{k}}{2}(g_{lk}^{\ast}+g_{nk})-\sum_{k,q}\omega_{k}\lambda_{q}g_{lk}^{\ast}g_{nk}(g_{lq}^{\ast}+g_{nq})\right)S_{ln}^{(g,g)}\right\}.
\end{eqnarray}
\end{widetext}
By calculating the deviation, we are capable to track the accuracy of the variational results.

\section{Equation of motion for the reduced effective density matrix\label{sec:appme} }

The effective density operator $\Lambda(t,t^{\prime})=U(t,t^{\prime})\sigma_{x}\rho(t^{\prime})U^{\dagger}(t,t^{\prime})$
satisfies the Liouville equation:
\begin{equation}
\frac{d}{dt}\Lambda(t,t^{\prime})=-i[H(t),\Lambda(t,t^{\prime})]
\end{equation}
with the initial condition $\Lambda(t^{\prime},t^{\prime})=\sigma_{x}\rho(t^{\prime}).$
The equation of motion can be transformed into the interaction picture,
yielding
\begin{equation}
\frac{d}{dt}\Lambda^{{\rm I}}(t,t^{\prime})=-i[H_{{\rm SR}}(t),\Lambda^{{\rm I}}(t,t^{\prime})]\equiv{\cal L}_{{\rm I}}(t)\Lambda^{{\rm I}}(t,t^{\prime}),\label{eq:LAI}
\end{equation}
where
\begin{equation}
\Lambda^{{\rm I}}(t,t^{\prime})=U_{{\rm {\rm S}}}^{\dagger}(t)\exp(iH_{{\rm R}}t)\Lambda(t,t^{\prime})U_{{\rm S}}(t)\exp(-iH_{{\rm R}}t),
\end{equation}
\begin{eqnarray}
H_{{\rm SR}}(t) & = & U_{{\rm S}}^{\dagger}(t)\exp(iH_{{\rm R}}t)H_{{\rm SR}}U_{{\rm S}}(t)\exp(-iH_{{\rm R}}t)\nonumber \\
 & = & \frac{\sigma_{x}(t)}{2}\sum_{k}\lambda_{k}(b_{k}e^{-i\omega_{k}t}+b_{k}^{\dagger}e^{i\omega_{k}t}),
\end{eqnarray}
are operators in the interaction picture.

Let the projection operator ${\cal P}$ be defined as
\begin{equation}
{\cal P}\rho={\rm Tr}_{{\rm R}}(\rho)\otimes\rho_{{\rm R}},
\end{equation}
where $\rho_{{\rm R}}$ is a fixed state of the reservoir. Let ${\cal Q}$
be the complementary projection operator such that
\begin{equation}
{\cal P}+{\cal Q}=I
\end{equation}
with $I$ the identity matrix. Accordingly, one finds that ${\cal P}^{2}={\cal P}$,
${\cal Q}^{2}={\cal Q}$, and ${\cal PQ}={\cal QP}=0$. With ${\cal P}$
and ${\cal Q}$, Eq. (\ref{eq:LAI}) can be partitioned into two parts:
\begin{eqnarray}
\frac{d}{dt}{\cal P}\Lambda^{{\rm I}}(t,t^{\prime}) & = & {\cal P}{\cal L}_{{\rm I}}(t)({\cal P}+{\cal Q})\Lambda^{{\rm I}}(t,t^{\prime}),\label{eq:dPLA}\\
\frac{d}{dt}{\cal Q}\Lambda^{{\rm I}}(t,t^{\prime}) & = & {\cal Q}{\cal L}_{{\rm I}}(t)({\cal P}+{\cal Q})\Lambda^{{\rm I}}(t,t^{\prime}).\label{eq:dQLA}
\end{eqnarray}
${\cal P}\Lambda^{{\rm I}}(t,t^{\prime})$ and ${\cal Q}\Lambda^{{\rm I}}(t,t^{\prime})$
are called the relevant part and irrelevant part, respectively. To
proceed, one solves the second equation and substitute its solution
into the first equation to derive a differential equation for the
relevant part. The second equation can be formally solved as
\begin{eqnarray}
{\cal Q}\Lambda^{{\rm I}}(t,t^{\prime}) & = & {\cal G}(t,t^{\prime}){\cal Q}\Lambda^{{\rm I}}(t^{\prime},t^{\prime})\nonumber \\
 &  & +\int_{t^{\prime}}^{t}ds{\cal G}(t,s){\cal Q}{\cal L}_{{\rm I}}(s){\cal P}\Lambda^{{\rm I}}(s,t^{\prime}),\label{eq:QLA}
\end{eqnarray}
where
\begin{equation}
{\cal G}(t,t^{\prime})={\cal T}_{\leftarrow}\exp\left[\int_{t^{\prime}}^{t}{\cal Q}{\cal L}_{{\rm I}}(s)ds\right]
\end{equation}
with ${\cal T}_{\leftarrow}$ being the time-ordering operator. The
operator $\Lambda^{{\rm I}}(s,t^{\prime})$ at time $s$ can be related
to $\Lambda^{{\rm I}}(t,t^{\prime})$ via

\begin{equation}
\Lambda^{{\rm I}}(s,t^{\prime})=G_{b}(t,s)\Lambda^{{\rm I}}(t,t^{\prime}),\label{eq:LAGLA}
\end{equation}
with $G_{b}(t,s)={\cal T}_{\rightarrow}\exp\left[-\int_{s}^{t}{\cal L}_{{\rm I}}(\tau)d\tau\right]$
being the backward unitary evolution operator. Substituting Eq. (\ref{eq:LAGLA})
into (\ref{eq:QLA}), one arrives at
\begin{eqnarray}
{\cal Q}\Lambda^{{\rm I}}(t,t^{\prime}) & = & {\cal G}(t,t^{\prime}){\cal Q}\Lambda^{{\rm I}}(t^{\prime},t^{\prime})+\int_{t^{\prime}}^{t}ds{\cal G}(t,s){\cal Q}{\cal L}_{{\rm I}}(s){\cal P}\nonumber \\
 &  & \times G_{b}(t,s)({\cal P}+{\cal Q})\Lambda^{{\rm I}}(t,t^{\prime}).\label{eq:QLAii}
\end{eqnarray}
Using
\begin{equation}
\Sigma(t,t^{\prime})=\int_{t^{\prime}}^{t}ds{\cal G}(t,s){\cal Q}{\cal L}_{{\rm I}}(s){\cal P}G_{b}(t,s),
\end{equation}
the irrelevant part ${\cal Q}\Lambda^{{\rm I}}(t,t^{\prime})$ can
be expressed as
\begin{eqnarray}
{\cal Q}\Lambda^{{\rm I}}(t,t^{\prime}) & = & \left[1-\Sigma(t,t^{\prime})\right]^{-1}\left[{\cal G}(t,t^{\prime}){\cal Q}\Lambda^{{\rm I}}(t^{\prime},t^{\prime})\right.\nonumber \\
 &  & \left.+\Sigma(t,t^{\prime}){\cal P}\Lambda^{{\rm I}}(t,t^{\prime})\right],\label{eq:QLAf}
\end{eqnarray}
where we used fact that $1-\Sigma(t,t^{\prime})$ can be inverted in a weak-coupling regime or at short times in strong coupling regimes~\cite{Breuer}. Substituting Eq. (\ref{eq:QLAf})
into (\ref{eq:dPLA}), we get the equation of motion for the relevant
part of $\Lambda(t,t^{\prime}):$
\begin{eqnarray}
\frac{d}{dt}{\cal P}\Lambda^{{\rm I}}(t,t^{\prime}) & = & {\cal P}{\cal L}_{{\rm I}}(t){\cal P}\Lambda^{{\rm I}}(t,t^{\prime})+{\cal I}(t,t^{\prime}){\cal Q}\Lambda^{{\rm I}}(t^{\prime},t^{\prime})\nonumber \\
 &  & +{\cal K}(t,t^{\prime}){\cal P}\Lambda^{{\rm I}}(t,t^{\prime}),
\end{eqnarray}
where
\begin{equation}
{\cal I}(t,t^{\prime})={\cal P}{\cal L}_{{\rm I}}(t)\left[1-\Sigma(t,t^{\prime})\right]^{-1}{\cal G}(t,t^{\prime}){\cal Q},
\end{equation}
\begin{equation}
{\cal K}(t,t^{\prime})={\cal P}{\cal L}_{{\rm I}}(t)\left[1-\Sigma(t,t^{\prime})\right]^{-1}\Sigma(t,t^{\prime}){\cal P}.
\end{equation}
The inhomogeneous part ${\cal I}(t,t^{\prime}){\cal Q}\Lambda^{{\rm I}}(t^{\prime},t^{\prime})$
can be simplified by using ${\cal Q}\Lambda^{{\rm I}}(t^{\prime},t^{\prime})=\sigma_{x}(t^{\prime}){\cal Q}\rho^{{\rm I}}(t^{\prime})$
and
\begin{equation}
{\cal Q}\rho^{{\rm I}}(t)=\left[1-\Sigma(t,t_{0})\right]^{-1}\Sigma(t,t_{0}){\cal P}\rho^{{\rm I}}(t).\label{eq:QrhoI}
\end{equation}
In deriving Eq. (\ref{eq:QrhoI}), we used the factorized initial
state $\rho^{{\rm I}}(t_{0})=\rho_{{\rm S}}^{{\rm I}}(t_{0})\otimes\rho_{{\rm R}}$
and the fact that ${\cal Q}\rho^{{\rm I}}(t)$ satisfies the same
equation as ${\cal Q}\Lambda^{{\rm I}}(t,t^{\prime})$. The equation
of motion becomes
\begin{eqnarray}
\frac{d}{dt}{\cal P}\Lambda^{{\rm I}}(t,t^{\prime}) & = & {\cal P}{\cal L}_{{\rm I}}(t){\cal P}\Lambda^{{\rm I}}(t,t^{\prime})+{\cal I}^{\prime}(t,t_{0}){\cal P}\rho^{{\rm I}}(t^{\prime})\nonumber \\
 &  & +{\cal K}(t,t^{\prime}){\cal P}\Lambda^{{\rm I}}(t,t^{\prime}),
\end{eqnarray}
where
\begin{eqnarray}
{\cal I}^{\prime}(t,t_{0}) & = & {\cal P}{\cal L}_{{\rm I}}(t)\left[1-\Sigma(t,t^{\prime})\right]^{-1}{\cal G}(t,t^{\prime}){\cal Q}\sigma_{x}(t^{\prime})\nonumber \\
 &  & \times\left[1-\Sigma(t^{\prime},t_{0})\right]^{-1}\Sigma(t^{\prime},t_{0}).
\end{eqnarray}
To proceed, we use the expansion $\left[1-\Sigma(t,t^{\prime})\right]^{-1}=\sum_{n=0}^{\infty}\left[\Sigma(t,t^{\prime})\right]^{n}$
and $\Sigma(t,t^{\prime})=\sum_{n=1}^{\infty}\Sigma_{n}(t,t^{\prime})$
($n$ indicates the order in the coupling strength $\lambda_k$). Up
to the second order in the coupling strength, and using ${\cal P}{\cal L}_{{\rm I}}(t){\cal P}=0$
(as we are interested in $\rho_{{\rm R}}=|\{0_{k}\}\rangle\langle\{0_{k}\}|$),
the kernels are given as
\begin{eqnarray}
{\cal {\cal I}^{\prime}}(t,t_{0}) & = & \int_{t_{0}}^{t^{\prime}}ds{\cal P}{\cal L}_{{\rm I}}(t)\sigma_{x}(t^{\prime}){\cal L}_{{\rm I}}(s){\cal P},
\end{eqnarray}

\begin{equation}
{\cal K}(t,t^{\prime})=\int_{t^{\prime}}^{t}ds{\cal P}{\cal L}_{{\rm I}}(t){\cal L}_{{\rm I}}(s){\cal P}.
\end{equation}
Finally, we obtain the second-order equation of motion
\begin{eqnarray}
\frac{d}{dt}{\cal P}\Lambda^{{\rm I}}(t,t^{\prime}) & = & \int_{t_{0}}^{t^{\prime}}ds{\cal P}{\cal L}_{{\rm I}}(t)\sigma_{x}(t^{\prime}){\cal L}_{{\rm I}}(s){\cal P}\rho^{{\rm I}}(t^{\prime})\nonumber \\
 &  & +\int_{t^{\prime}}^{t}ds{\cal P}{\cal L}_{{\rm I}}(t){\cal L}_{{\rm I}}(s){\cal P}\Lambda^{{\rm I}}(t,t^{\prime}).\label{eq:dPLA2}
\end{eqnarray}
From Eq. (\ref{eq:dPLA2}), one readily derives Eq. (\ref{eq:correq})
in the Schr\"{o}dinger picture.

Similarly to $\Lambda_{{\rm S}}(t,t^{\prime}),$the reduced density
matrix $\rho_{{\rm S}}(t)$ is also obtained via the projection method.
To arrive at Eq. (\ref{eq:me}) , we use the factorized initial state
$\rho(0)=\rho_{{\rm S}}(0)\otimes|\{0_{k}\}\rangle\langle\{0_{k}\}|$.

\section{Equations of motion in the Floquet picture}\label{sec:appFP}
We numerically solve Eqs.~(\ref{eq:correq}) and~(\ref{eq:me}) with
the aid of Floquet theory~\cite{shirley,grifoni}, which states that the evolution operator
of the driven qubit takes the form:
\begin{equation}
U_{{\rm S}}(t)=\sum_{\gamma=1}^{2}|u_{\gamma}(t)\rangle\langle u_{\gamma}(0)|e^{-i\varepsilon_{\gamma}t},
\end{equation}
where $|u_{\gamma}(t)\rangle=|u_{\gamma}(t+2\pi/\omega_{x})\rangle$
is the Floquet state with the real quasienergy $\varepsilon_{\gamma}.$
It is straightforward to show that $|u_{\gamma}(t)\rangle$ and $\varepsilon_{\gamma}$
satisfy the following equation:
\begin{equation}
[H_{{\rm S}}(t)-i\partial_{t}]|u_{\gamma}(t)\rangle=\varepsilon_{\gamma}|u_{\gamma}(t)\rangle.
\end{equation}
By employing the Sambe space, this differential equation can be solved numerically to yield the Floquet states
and quasienergies~\cite{shirley,grifoni}. In terms
of the Floquet states, we define the following matrix elements and decay rate:
\begin{equation}
\Lambda_{\mu\nu}(t,t^{\prime})=\langle u_{\mu}(t)|\Lambda_{{\rm S}}(t,t^{\prime})|u_{\nu}(t)\rangle,
\end{equation}
\begin{equation}
\rho_{\mu\nu}(t)=\langle u_{\mu}(t)|\rho_{{\rm S}}(t)|u_{\nu}(t)\rangle,
\end{equation}
\begin{equation}
X_{\mu\nu}(t)=\langle u_{\mu}(t)|\sigma_{x}|u_{\nu}(t)\rangle,
\end{equation}
\begin{equation}
X_{\mu\nu,n}=\frac{\omega_{x}}{2\pi}\int_{0}^{2\pi/\omega_{x}}X_{\mu\nu}(t)e^{-in\omega_{x}t}dt,
\end{equation}
\begin{equation}
\Gamma(\omega,t,t^{\prime})=\int_{t^{\prime}}^{t}C(\tau)e^{-i\omega\tau}d\tau.
\end{equation}
With these quantities, we can rewrite the equation of motion for the
effective density operator as \begin{widetext}
\begin{eqnarray}
\frac{d}{dt}\Lambda_{{\rm \mu\nu}}(t,t^{\prime}) & = & -i\Delta_{\mu\nu}\Lambda_{{\rm \mu\nu}}(t,t^{\prime})-\sum_{\gamma,\delta}{\cal K}_{\mu\nu,\gamma\delta}(t-t^{\prime},0)\Lambda_{\gamma\delta}(t,t^{\prime})-{\cal I}_{\mu\nu}(t,t^{\prime}),
\end{eqnarray}
where
\begin{equation}
\Delta_{\mu\nu}=\varepsilon_{\mu}-\varepsilon_{\nu},
\end{equation}
\begin{eqnarray}
{\cal K}_{\mu\nu,\gamma\delta}(t,t^{\prime}) & = & \sum_{n}e^{in\omega_{x}t}\left\{ \sum_{\lambda}\delta_{\nu,\delta}X_{\mu\lambda}(t)X_{\lambda\gamma,n}\Gamma(\Delta_{\lambda\gamma,n},t,t^{\prime})-X_{\mu\gamma,n}X_{\delta\nu}(t)\Gamma(\Delta_{\mu\gamma,n},t,t^{\prime})\right.\nonumber \\
 &  & \left.+\sum_{\lambda}\delta_{\mu,\gamma}X_{\delta\lambda,n}X_{\lambda\nu}(t)\Gamma^\ast(-\Delta_{\delta\lambda,n},t,t^{\prime})-X_{\mu\gamma}(t)X_{\delta\nu,n}\Gamma^{\ast}(-\Delta_{\delta\nu,n},t,t^{\prime})\right\} ,
\end{eqnarray}
and ${\cal I}_{\mu\nu}(t,t^{\prime})=\langle u_{\mu}(t)|{\cal I}(t,t^{\prime})|u_{\nu}(t)\rangle$
is the element of the following matrix:
\begin{equation}
{\cal I}(t,t^{\prime})=[X(t),\sigma_{x}(t,t^{\prime})F(t,t^{\prime})\rho_{{\rm S}}(t,t^{\prime})]+[\sigma_{x}(t,t^{\prime})\rho_{{\rm S}}(t,t^{\prime})F^{\dagger}(t,t^{\prime}),X(t)],
\end{equation}
\end{widetext}where
\begin{equation}
X(t)=\sum_{\mu,\nu}|u_{\mu}(t)\rangle\langle u_{\nu}(t)|X_{\mu\nu}(t),
\end{equation}
\begin{equation}
\sigma_{x}(t,t^{\prime})=\sum_{\mu,\nu}|u_{\mu}(t)\rangle\langle u_{\nu}(t)|X_{\mu\nu}(t^{\prime})\exp(-i\Delta_{\mu\nu}t),
\end{equation}
\begin{equation}
\rho_{{\rm S}}(t,t^{\prime})=\sum_{\mu,\nu}|u_{\mu}(t)\rangle\langle u_{\nu}(t)|\rho_{\mu\nu}(t^{\prime})\exp(-i\Delta_{\mu\nu}t),
\end{equation}
\begin{equation}
F(t,t^{\prime})=\sum_{\mu,\nu}|u_{\mu}(t)\rangle\langle u_{\nu}(t)|\sum_{n}e^{in\omega_{x}t}X_{\mu\nu,n}\Gamma(\Delta_{\mu\nu,n},t,t-t^{\prime}).
\end{equation}
Similarly, $\rho_{\mu\nu}(t^{\prime})$ is also calculated in the
Floquet picture.

\section{Hierarchy equations of motion}\label{sec:heom}
Let us denote the eigenstates for $\sigma_z$ as $\sigma$, then the reduced density matrix element for the two-level system is expressed in the path integral form with the factorized initial condition as \cite{PI1,PI2,PI3}
\begin{eqnarray}\label{PIrho}
\rho(\sigma,\sigma^{\prime};t)&=&\int\mathcal{D}\sigma\int\mathcal{D}\sigma^{\prime}\rho(\sigma_0,\sigma_0^{\prime}; t_0)\nonumber\\
& &\times e^{iS[\sigma;t]}F(\sigma,\sigma^{\prime};t)e^{-iS[\sigma^{\prime};t]}.
\end{eqnarray}
Here, $S[\sigma;t]$ is the action of the two-level system, and $F(\sigma,\sigma^{\prime};t)$ is the Feynman-Vernon influence functional, given by
\begin{widetext}
\begin{equation}
F(\sigma,\sigma^{\prime};t)=\exp\left(-\int_0^{\infty}d\omega{J(\omega)}\int_{t_0}^td\tau\int_{t_0}^\tau{d\tau^{\prime}}V^{\times}(\tau)\left[V^{\times}(\tau^{\prime})\coth(\frac{\beta\omega}{2})\cos(\omega(\tau-\tau^{\prime}))-iV^{\circ}(\tau^{\prime})\sin(\omega(\tau-\tau^{\prime}))\right]\right).
\end{equation}
\end{widetext}
Here, we have introduced the abbreviations:
\begin{eqnarray}
V&=&\frac{\sigma_x}{2},\\
V^{\times}&=&V[\tau]-V[\tau^{\prime}],\\
V^{\circ}&=&V[\tau]+V[\tau^{\prime}].
\end{eqnarray}
The correlation function can be written as
\begin{equation}\label{CT}
C(t)=\int_0^{\infty}d\omega{J}(\omega)\left[\coth(\frac{\beta\omega}{2})\cos(\omega{t})-i\sin(\omega{t})\right],
\end{equation}
where $\beta$ is the inverse of the temperature. For the zero temperature case considered in this work, we have
\begin{eqnarray}\label{CTZero}
C(t)&=&2\alpha\left[\frac{\cos(\omega_ct)-1}{t^2}+\frac{\omega_c\sin(\omega_ct)}{t}\right]\nonumber\\
& &-2i\alpha\left[\frac{\sin(\omega_ct)}{t^2}-\frac{\omega_c\cos(\omega_c t)}{t}\right]\nonumber \\
&\equiv&C_R(t)+iC_I(t).
\end{eqnarray}
where $C_R(t)$ and $C_I(t)$ are real and imaginary parts of the correlation function, respectively. We apply an incomplete set of oscillatory exponentially decaying functions (OEDFs) for an approximate decomposition~\cite{BCD},
\begin{equation}\label{CFit}
C_X^{\mathrm{fit}}(t)=\sum_{n=1}^{N_X}a_{X;2n-1}\cos(\omega_{X;n}t)e^{-\gamma_{X;n}t}+a_{X;2n}\sin(\omega_{X;n}t)e^{-\gamma_{X;n}t}.
\end{equation}
The fitting parameters, $\left\lbrace{a}_{X;n},\omega_{X;n},\gamma_{X;n}\right\rbrace$, are allowed to be uncorrelated for the real ($X=R$) and imaginary ($X=I$) parts.

We can thus express the correlation function~(\ref{CTZero}) as
\begin{equation}
C(t)\approx\sum_{n=1}^{\mathrm{2N_R}}a_{R;n}\varphi_{R;n}(t)+i\sum_{m=1}^{\mathrm{2N_I}}a_{I;m}\varphi_{I;m}(t).
\end{equation}
The basis functions are OEDFs, given by
\begin{widetext}
\begin{equation}
\left\lbrace\varphi_{X;n}(t)\right\rbrace=\left\lbrace\cos(\omega_{X;n}{t})e^{-\gamma_{X;n}t},\sin(\omega_{X;n}{t})e^{-\gamma_{X;n}{t}},n=1,\cdots,N_{X}\right\rbrace,
\end{equation}
with $X=R$ and $I$. For the two separated basis functions, $\left\lbrace\varphi_{R;n}(t)\right\rbrace$ and $\left\lbrace\varphi_{I;m}(t)\right\rbrace$, we have relations, $\partial_t\varphi_{R;n}(t)=\sum_{n'}\eta_{R;n,n'}\varphi_{R;n'}(t)$ and $\partial_t\varphi_{I;m}(t)=\sum_{m'}\eta_{I;m,m'}\varphi_{I;m'}(t)$, where
\begin{equation}
\eta_{X,n,n'}=\left(\begin{array}{ccccccc}
-\gamma_{X;1} & -\omega_{X;1} & 0 & 0 & \cdots & 0 & 0 \\
\omega_{X;1} & -\gamma_{X;1} & 0 & 0 & \cdots & 0  & 0\\
0 & 0 & -\gamma_{X;2} & -\omega_{X;2} & \cdots & 0& 0 \\
0 & 0 & \omega_{X;2} & -\gamma_{X;2} & \cdots & 0 & 0 \\
\vdots & \vdots & \vdots & \vdots & \ddots & \vdots  & \vdots\\
0 & 0 & 0 & 0 & \cdots & -\gamma_{X;N_X} &-\omega_{X;N_X} \\
0 & 0 & 0 & 0 & \cdots & \omega_{X;N_X} & -\omega_{X;N_X} \\
\end{array}
\right).
\end{equation}

The influence functional equation can be expressed as
\begin{eqnarray}
F(\sigma,\sigma^{\prime};t)&=&\prod_{n=1}^{2N_R}\exp\left(-\int_{t_0}^td\tau\int_{t_0}^{\tau}d\tau^{\prime}V^{\times}(\tau)V^{\times}(\tau^{\prime})a_{R;n}\varphi_{R;n}(\tau-\tau^{\prime})\right)\nonumber \\
&&\times\prod_{m=1}^{2N_I}\exp\left(-\int_{t_0}^td\tau\int_{t_0}^{\tau}d\tau^{\prime}V^{\times}(\tau)V^{\circ}(\tau^{\prime})ia_{I;m}\varphi_{I;m}(\tau-\tau^{\prime})\right).
\end{eqnarray}
Taking the derivative of Eq.~(\ref{PIrho}), we have
\begin{eqnarray}
\frac{\partial}{\partial{t}}\rho(\sigma,\sigma^{\prime};t)&=&-i\mathcal{L}\rho(\sigma,\sigma^{\prime};t)-V^{\times}(t)\int\mathcal{D}\sigma\int\mathcal{D}\sigma^{\prime}\rho(\sigma_0,\sigma_0^{\prime};t_0)\nonumber \\
&&\times\left[\int_{t_0}^td\tau{V}^{\times}(\tau)\sum_{n=1}^{2N_R}a_{R;n}\varphi_{R;n}(t-\tau)+\int_{t_0}^{t}d\tau{V}^{\circ}(\tau)\sum_{m=1}^{2N_I}ia_{I;m}\varphi_{I;m}(t-\tau)\right]\nonumber \\
&&\times{e}^{iS[\sigma,t]}F(\sigma,\sigma^{\prime};t)e^{-iS[\sigma^{\prime};t]},
\end{eqnarray}
where ${\cal L}$ is the Liouville superoperator describing the unitary evolution governed by $H_{\rm S}(t)$.
In order to derive the equation of motion, we introduce the auxiliary operator $\rho_{j_1,\cdots,j_{2N_R};k_1,\cdots,k_{2N_I}}$ by its matrix element as
\begin{eqnarray}
\rho_{j_1,\cdots,j_{2N_R};k_1,\cdots,k_{2N_I}}(\sigma,\sigma^{\prime};t)=&&\int\mathcal{D}\sigma\int\mathcal{D}\sigma^{\prime}\rho(\sigma_0,\sigma_0^{\prime};t_0)\prod_{n=1}^{2N_R}\left(\int_{t_0}^td\tau{V}^{\times}(\tau)\varphi_{R;n}(t-\tau)\right)^{j_n}\nonumber \\
&&\times\prod_{m=1}^{2N_I}\left(\int_{t_0}^td\tau{V}^{\circ}(\tau)\varphi_{I;m}(t-\tau)\right)^{k_m}e^{iS[\sigma,t]}F(\sigma,\sigma^{\prime};t)e^{-iS[\sigma^{\prime};t]}
\end{eqnarray}
for non-negative integers $j_1,\cdots,j_{2N_R};k_1,\cdots,k_{2N_I}$. It should be noted that $\rho_{0,\cdots,0}(t)=\rho(t)$ denote the true reduced density matrix, while other auxiliary density matrices are introduced to take into account all orders of system-bath couplings. Differentiating $\rho_{j_1,\cdots,j_{2N_R};k_1,\cdots,k_{2N_I}}(\sigma,\sigma^{\prime};t)$ with respect to $t$, we obtain the following hierarchy of equations in operator form:
\begin{eqnarray}\label{HEOM}
&&\partial_t\rho_{j_1,\cdots,j_{2N_R};k_1,\cdots,k_{2N_I}}(\sigma,\sigma^{\prime};t)\nonumber \\
=&&-i\mathcal{L}\rho_{j_1,\cdots,j_{2N_R};k_1,\cdots,k_{2N_I}}(\sigma,\sigma^{\prime};t)\nonumber \\
&&+V^{\times}(t)\sum_{n=1}^{2N_R}j_n\varphi_{R;n}(0)\rho_{j_1,\cdots, j_n-1,\cdots,j_{2N_R};k_1,\cdots,k_{2N_I}}(\sigma,\sigma^{\prime};t)\nonumber \\
&&+V^{\circ}(t)\sum_{m=1}^{2N_I}k_m\varphi_{I;m}(0)\rho_{j_1,\cdots,j_{2N_R};k_1,\cdots,k_m-1,\cdots,k_{2N_I}}(\sigma,\sigma^{\prime};t)\nonumber \\
&&+\sum_{n=1}^{2N_R}\sum_{n'=1}^{2N_R}j_n\eta_{R;n,n'}\rho_{j_1,\cdots,j_n-1,\cdots,j_{n'}+1,\cdots,j_{2N_R};k_1,\cdots,k_{2N_I}}(\sigma,\sigma^{\prime};t)\nonumber \\
&&+\sum_{m=1}^{2N_I}\sum_{m'=1}^{2N_I}k_m\eta_{I;m,m'}\rho_{j_1,\cdots,j_{2N_R};k_1,\cdots,k_m-1,\cdots,k_{m'}+1,\cdots,k_{2N_I}}(\sigma,\sigma^{\prime};t)\nonumber \\
&&-V^{\times}(t)\sum_{n=1}^{2N_R}a_{R;n}\rho_{j_1,\cdots, j_n+1,\cdots,j_{2N_R};k_1,\cdots,k_{2N_I}}(\sigma,\sigma^{\prime};t)\nonumber \\
&&-V^{\times}(t)\sum_{m=1}^{2N_I}ia_{I;m}\rho_{j_1,\cdots,j_{2N_R};k_1,\cdots,k_m+1,\cdots,k_{2N_I}}(\sigma,\sigma^{\prime};t).
\end{eqnarray}

The HEOM consists of an infinite number of equations, which must be truncated for practical simulations. For this purpose, the integers $j_1,\cdots,j_{2N_R};k_1,\cdots,k_{2N_I}$ should satisfy $\sum_{n=1}^{2N_R}j_n+\sum_{m=1}^{2N_I}k_m\leq\mathrm{N}_\mathrm{trun}$, where $\mathrm{N}_\mathrm{trun}$ is the depth of the hierarchy.
\end{widetext}

\end{document}